\documentclass[twocolumn,showpacs]{revtex4}
\usepackage{graphics}

\begin{document}

\title{Properties of entangled photon pairs generated in one-dimensional nonlinear
photonic-band-gap structures}

\author{Jan Pe\v{r}ina, Jr.}
\affiliation{Joint Laboratory of Optics of Palack\'{y} University
and Institute of Physics of Academy of Sciences of the Czech
Republic, 17. listopadu 50A, 772 07 Olomouc, Czech Republic}
\email{perina_j@sloup.upol.cz}
\author{Marco Centini}
\author{Concita Sibilia}
\author{Mario Bertolotti}
\affiliation{Dipartimento di Energetica, Universit\`{a} La
Sapienza di Roma, Via A. Scarpa 16, 00161 Roma, Italy}
\author{Michael Scalora}
\affiliation{Charles M. Bowden Research Center, RD\&EC, Redstone
Arsenal, Bldg 7804, Alabama 35898-5000}

\begin{abstract}
We have developed a rigorous quantum model of spontaneous
parametric down-conversion in a nonlinear 1D photonic-band-gap
structure based upon expansion of the field into monochromatic
plane waves. The model provides a two-photon amplitude of a
created photon pair. The spectra of the signal and idler fields,
their intensity profiles in the time domain, as well as the
coincidence-count interference pattern in a Hong-Ou-Mandel
interferometer are determined both for cw and pulsed pumping
regimes in terms of the two-photon amplitude. A broad range of
parameters characterizing the emitted down-converted fields can be
used. As an example, a structure composed of 49 layers of GaN/AlN
is analyzed as a suitable source of photon pairs having high
efficiency.
\end{abstract}



\maketitle

\section{Introduction}

More than 20 years ago, Hong, Ou, and Mandel showed experimentally
that mutually strongly quantum correlated (entangled) photon pairs
can be emitted in the nonlinear process of parametric
down-conversion \cite{Hong1987,Mandel1995} at the single-photon
level. For the occurrence of entangled photon pairs, the
spontaneous character of the process is important. Entanglement of
two photons comprising a photon pair might occur for various
physical quantities like frequencies, emission angles
(wave-vectors), or polarizations. Perhaps most interestingly,
entangled photon pairs manifest themselves in time domain, where
both photons are detected within a relatively narrow time window.
This is a direct consequence of the `point' character of the
emission of a photon pair in the time domain.  The width of the
time window characterized by an entanglement time is typically on
the order of several hundreds of fs, and is experimentally
observable using Hong-Ou-Mandel interferometer.

In the time that has intervened since the original predictions,
entangled photon pairs have been used for numerous experiments
demonstrating both fundamental aspects of their physical
properties \cite{Perina1994} and their potential for applications.
These also include tests of Bell inequalities \cite{Perina1994},
quantum teleportation \cite{Bouwmeester1997}, the generation of
Greenberger-Horne-Zeilinger states \cite{Bouwmeester1999} or
quantum computation \cite{Bouwmeester2000} that can exploit photon
pairs. Quantum cryptography with photon pairs
\cite{Lutkenhaus2000} may be considered the most important
application. We mention that applications to metrology have also
been suggested \cite{Migdal1999}.

Most researchers in the field have focused their attention for
the most part on  bulk nonlinear crystals, pumped by intense
laser beams in type I and type II
configurations. Bright sources of polarization-entangled photon
pairs have been fabricated using two nonlinear type I crystals, mutually rotated
by 90 degrees \cite{Kwiat1999,Kwiat2001,Nambu2002}. The nonlinear
processes in an optical cavity have also been used to enhance the
photon-pair generation rate \cite{Shapiro2000}. In addition,
periodically-poled materials may be used to increase the
photon-pair generation rate in materials where phase matching is not
naturally available \cite{Kuklewicz2005}. At present, even
sources of photon pairs pumped by laser diodes have been developed
\cite{Trojek2004}.

On the fabrication front, techniques of structures composed of
nonlinear thin layers (of width of several tens or hundreds of nm)
have developed to the point that useful structures with
pre-defined properties may be achieved rather easily
\cite{Bertolotti2001}. These nonlinear photonic-band-gap
structures
\cite{Yablonovitch1987,John1987,Joannopoulos1995,Sakoda2005} are
very promising as sources of photon pairs, as has recently been
shown in \cite{Vamivakas2004,Centini2005}. Despite the small
amount of nonlinear material embedded inside them, they can
generate photon pairs with relatively high efficiency thanks to
the constructive interference that involves both the pump and the
down-converted fields, their spatial inhomogeneity
nothwithstanding. The enhancement of the photon-pair generation
rate has been predicted to be several hundreds and even thousands
of times larger than photon-pair generation rates in nonlinear,
bulk materials \cite{Centini2005}. Moreover, the spectral and
spatial characteristics of the down-converted fields depend on
details of the structure, a fact that might be used to control the
process, at least to some extent. For example, photon pairs with
very narrow spectra may be obtained from suitable structures. We
note that along the same vein, four-wave mixing in
photonic-band-gap nonlinear fibers is also promising as a modern
source of photon pairs \cite{Li2005,Fulconis2005}.

In this paper, we present a quantum model of photon-pair emission
in a nonlinear, one-dimensional photonic-band-gap  structure, based
upon a perturbative solution of the Schr\"{o}dinger equation. This model
extends those developed for bulk nonlinear materials in
\cite{Keller1997,Perinajr1999,DiGiuseppe1997,Grice1998}. It has
been shown in \cite{Centini2005} that this approach is compatible
with models based on methods of classical nonlinear optics
(multiple-scale spatial and temporal expansion methods) with a specific kind of
stochastic averaging concerning emitted spectra. The quantum
model, however, provides a complete description of the nonlinear
process.

The paper is organized as follows. In Sec. II, the model
is presented in three steps. Description of a quantum field in a
layered medium in Subsec. IIA is followed by the description of
nonlinear quantum interactions in Subsec. IIB. In Subsec. IIC
measurable characteristics of the generated down-converted fields
are determined. The behavior of physical quantities characterizing a
photon pair are discussed in Sec. III, both for cw and pulsed
pumping regimes. Sec. IV contains our conclusions.

\section{Description of spontaneous parametric down-conversion
in a one-dimensional, nonlinear photonic-band-gap structure}

\begin{figure}    
 \resizebox{0.8\hsize}{!}{\includegraphics{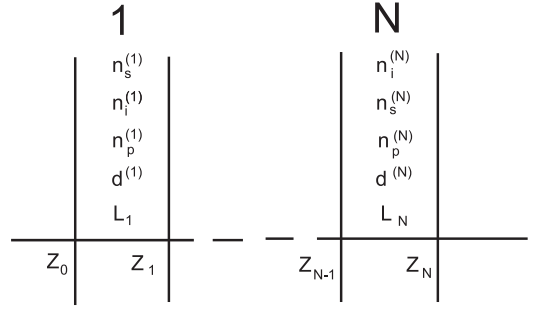}}
 \vspace{2mm}
 \caption{Scheme of the considered nonlinear one-dimensional layered structure:
 $ n_s^{(l)} $, $ n_i^{(l)} $, and $ n_p^{(l)} $ denote indices of refraction of
 signal, idler, and pump fields in $ l $th layer with the length $ L_l $,
 $ d^{(l)} $ stands for a nonlinear coefficient of $ l $th layer
 and $ z_l $ are positions of the boundaries.}
 \label{fig1}
\end{figure}

We consider a stack of $ N $ nonlinear layers in which spontaneous
parametric down-conversion may occur. A sketch of the system is
shown in Fig.~\ref{fig1}. The $ l $th layer begins at $ z=z_{l-1}
$ and ends at $ z=z_l $, its length is denoted as $ L_l $, $
l=1,\ldots, N $. Linear indices of refraction of pump, signal, and
idler fields in the $ l $th layer are denoted as $ n_p^{(l)} $, $
n_s^{(l)} $, and $ n_i^{(l)} $, respectively. Symbols $ n_s^{(0)}
$, $ n_i^{(0)} $, and $ n_p^{(0)} $ ($ n_s^{(N+1)} $, $
n_i^{(N+1)} $, and $ n_p^{(N+1)} $) mean indices of refraction in
front of (beyond) the sample. The symbol $ d^{(l)} $ is used for
the nonlinear tensor of $ l $th layer; symbols $ k_p^{(l)} $, $
k_s^{(l)} $, and $ k_i^{(l)} $ stand for pump-, signal-, and
idler- field wave-vectors in the $ l $th layer.

\subsection{Optical fields in a one-dimensional photonic-band-gap structure}

\begin{figure}    
 \resizebox{0.6\hsize}{!}{\includegraphics{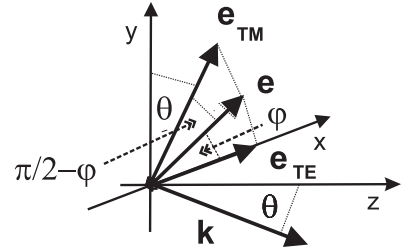}}
 \vspace{2mm}
 \caption{Scheme of the used coordinate system for the description of
 pump, signal, and idler fields. A field with wave-vector $ {\bf
 k} $ (lying in $ yz $ plane) and polarization vector $ {\bf e} $ (perpendicular
 to $ {\bf k} $) propagates at an angle $ \vartheta $ with
 respect to $ z $ axis. The field is linearly polarized and the polarization vector
 $ {\bf e} $ makes an angle $ \varphi $ with respect to $ x $ axis.
 The polarization vector $ {\bf e}_{TE} $ of TE wave is parallel to $ x $ axis whereas
 the polarization vector $ {\bf e}_{TM} $ of TM wave lyes in $ yz
 $ plane.}
 \label{fig2}
\end{figure}

The structure is pumped by a classical strong pump field that
propagates under the angle $ \vartheta_p $ with respect to $ z $
axis (see Fig.~\ref{fig2}). Its wave-vector $ {\bf k}_p $ lyes in
the $ yz $ plane, whereas its electric-field amplitude $ {\bf
E}_{p_F} $ is perpendicular to the wave-vector $ {\bf k}_p $.
The pump-field positive-frequency electric-field amplitude at the
input of the structure at $ z=z_0 $ is denoted as $ {\bf
E}_{p_F}^{(+)}(z_0,t) $ and can be conveniently described using a
positive-frequency electric-field amplitude spectrum $ {\bf
E}_{p_F}^{(+)}(z_0,\omega_p) $ determined as follows:
\begin{equation}    
 {\bf E}_{p_F}^{(+)}(z_0,\omega_p) = \frac{1}{\sqrt{2\pi}} \int_{-\infty}^{\infty} dt \,
 {\bf E}_{p_F}^{(+)}(z_0,t)\exp(i\omega_p t).
\label{1}
\end{equation}
If we assume a Gaussian time profile and a linear polarization of
the incident electric-field amplitude $ {\bf E}_{p_F} $ in the
direction rotated by an angle $ \varphi_p $ with respect to an
incident TE-wave polarization direction $ {\bf e}_{p_F,TE} $ ($
{\bf e}_{p_F,TM} $ denotes an incident TM-wave polarization
direction, see Fig.~\ref{fig2}) we have:
\begin{eqnarray}   
 {\bf E}_{p_F}^{(+)}(z_0,t) &=& \left[ {\bf e}_{p_F,TE} \cos(\varphi_p) +
 {\bf e}_{p_F,TM} \sin(\varphi_p) \right] \nonumber \\
 & & \mbox{} \times
 \xi_p \exp \left( - \frac{1+ia_p}{\tau_p^2}t^2 \right)
 \exp(-i\omega_p^0 t), \nonumber \\
 &=& {\bf E}_{p_F,TE}^{(+)}(z_0,t) + {\bf
 E}_{p_F,TM}^{(+)}(z_0,t),
\label{2}
\end{eqnarray}
where $ \xi_p $ is the pump-pulse amplitude, $ \tau_p $ pulse
duration, $ \omega_p^0 $ central frequency, and $ a_p $ denotes a
chirp parameter of the pulse. The spectrum $ {\bf
E}_{p_F}^{(+)}(z_0,\omega_p) $ determined by Eq. (\ref{1}) is
given as:
\begin{eqnarray}   
 {\bf E}_{p_F}^{(+)}(z_0,\omega_p) &=&  \left[ {\bf e}_{p_F,TE} \cos(\varphi_p) +
 {\bf e}_{p_F,TM} \sin(\varphi_p) \right] \nonumber \\
 & & \mbox{} \hspace{-2.5cm} \times \xi_p \frac{\tau_p}{\sqrt{2(1+ia_p)}}
  \exp \left( - \frac{\tau_p^2}{4(1+ia_p)} (\omega_p-\omega_p^0)^2 \right)
  \nonumber \\
 &=& {\bf E}_{p_F,TE}^{(+)}(z_0,\omega_p) + {\bf
 E}_{p_F,TM}^{(+)}(z_0,\omega_p).
\end{eqnarray}
On the other hand, the following spectrum corresponds to cw
pumping:
\begin{eqnarray}   
 {\bf E}_{p_F}^{(+)}(z_0,\omega_p) &=& \left[ {\bf e}_{p_F,TE} \cos(\varphi_p) +
 {\bf e}_{p_F,TM} \sin(\varphi_p) \right] \nonumber \\
 & & \mbox{} \times \xi_p \delta(\omega_p-\omega_p^0);
\label{4}
\end{eqnarray}
$ \delta $ means a Dirac delta function.

The pump field incident on the structure is
scattered at each boundary inside the  structure, so as to achieve a certain profile
along the $ z $ axis, provided that the nonlinear interaction does not lead to
pump-field depletion. Scattering of the pump field is
conveniently described using its decomposition into monochromatic
waves. The positive-frequency electric-field amplitude $ {\bf
E}_{p,\alpha}^{(+)}(z,\omega_p) $ of a monochromatic component at
frequency $ \omega_p $ with polarization $ \alpha $ ($ \alpha = $
TE, TM) can be written as follows \cite{Yeh1988}:
\begin{eqnarray}     
 {\bf E}_{p,\alpha}^{(+)}(z,\omega_p) &=& {\rm rect}_{-\infty,z_0}(z)
   \nonumber \\
  & & \mbox{} \hspace{-2cm} \times \left[
   A_{p_F,\alpha}^{(0)}(\omega_p){\bf
   e}_{p_F,\alpha}^{(0)}(\omega_p)
  \exp(i{\bf k}_{p,z}^{(0)}(z-z_0)) \right. \nonumber \\
  & & \mbox{} \hspace{-2cm} \left. + A_{p_B,\alpha}^{(0)}(\omega_p)
   {\bf e}_{p_B,\alpha}^{(0)}(\omega_p)
   \exp(-i{\bf k}_{p,z}^{(0)}(z-z_0)) \right]
   \nonumber \\
  & & \mbox{} \hspace{-2cm}+ \sum_{l=1}^{N} {\rm rect}_{z_{l-1},z_l}(z) \nonumber \\
  & & \mbox{} \hspace{-2cm} \times \left[
   A_{p_F,\alpha}^{(l)}(\omega_p){\bf e}_{p_F,\alpha}^{(l)}(\omega_p)
   \exp(i{\bf k}_{p,z}^{(l)}(z-z_{l-1})) \right. \nonumber \\
  & & \mbox{} \hspace{-2cm}\left. + A_{p_B,\alpha}^{(l)}(\omega_p)
   {\bf e}_{p_B,\alpha}^{(l)}(\omega_p)
   \exp(-i{\bf k}_{p,z}^{(l)}(z-z_{l-1})) \right]
   \nonumber \\
  & & \mbox{} \hspace{-2cm}+ {\rm rect}_{z_N,\infty}(z) \nonumber \\
  & & \mbox{} \hspace{-2cm}\times \left[
   A_{p_F,\alpha}^{(N+1)}(\omega_p) {\bf e}_{p_F,\alpha}^{(N+1)}(\omega_p)
   \exp(i{\bf k}_{p,z}^{(N+1)}(z-z_N)) \right. \nonumber \\
  & & \mbox{} \hspace{-2cm}\left. +
  A_{p_B,\alpha}^{(N+1)}(\omega_p)
   {\bf e}_{p_B,\alpha}^{(N+1)}(\omega_p)
   \exp(-i{\bf k}_{p,z}^{(N+1)}(z-z_N)) \right];
   \nonumber \\
  & & \alpha=TE,TM;
\label{5}
\end{eqnarray}
the function $ {\rm rect}_{z_a,z_b}(z) $ equals one for $ z_a \le
z \le z_b $ and is zero otherwise. Polarization vectors of $
\alpha $-waves in $ l $th layer are denoted as $ {\bf
e}^{(l)}_{p_F,\alpha} $ and $ {\bf e}^{(l)}_{p_B,\alpha} $ for
forward- and backward-propagating fields with respect to $ z $
axis, respectively, and they are frequency dependent. The symbol $
{\bf k}_{p,z}^{(l)} $ denotes a $ z $ component of the pump-field
wave-vector $ {\bf k}_{p}^{(l)} $ in $ l $th layer and is
determined by the expression
\begin{equation}    
 {\bf k}_{p,z}^{(l)} = {\bf k}_{p}^{(l)} \cos(\vartheta_p^{(l)}),
\end{equation}
where the wave-vector $ {\bf k}_{p}^{(l)} $ propagates in the $ l
$th layer under the angle $ \vartheta_p^{(l)} $ with respect to $
z $ axis. The angles $ \vartheta_p^{(l)} $ fulfill the Snell law
at the boundaries, i.e.
\begin{equation}  
 n_p^{(l)}\sin(\vartheta_p^{(l)}) =
 n_p^{(l+1)}\sin(\vartheta_p^{(l+1)}),
  \hspace{5mm} l=0,\ldots,N;
\end{equation}
$ \vartheta_p^{(0)} = \vartheta_p $.

The symbol  $ A_{p_F,\alpha}^{(0)}(\omega_p) $ is identified as an
electric-field amplitude of the pump field at frequency $ \omega_p
$ with polarization $ \alpha $ incident on the structure from the
left-hand side, whereas the symbol $
A_{p_B,\alpha}^{(N+1)}(\omega_p) $ describes an electric-field
amplitude of the pump field at frequency $ \omega_p $ with
polarization $ \alpha $ incident from the right-hand side. The
remaining amplitudes $ A_{p_F,\alpha}^{(l)}(\omega_p) $ and $
A_{p_B,\alpha}^{(l)}(\omega_p) $ are determined using relations at
the boundaries and free-field propagation inside the layers:
\begin{eqnarray}  
 \pmatrix{ A_{p_F,\alpha}^{(1)}(\omega_p) \cr A_{p_B,\alpha}^{(1)}(\omega_p)} &=&
   {\cal T}_{p,\alpha}^{(0)}(\omega_p)\pmatrix{ A_{p_F,\alpha}^{(0)}(\omega_p) \cr
   A_{p_B,\alpha}^{(0)}(\omega_p) }, \nonumber \\
 \pmatrix{ A_{p_F,\alpha}^{(l+1)}(\omega_p) \cr A_{p_B,\alpha}^{(l+1)}(\omega_p)
   } &=& {\cal T}_{p,\alpha}^{(l)}(\omega_p) {\cal P}_{p}^{(l)}(\omega_p)
   \pmatrix{ A_{p_F,\alpha}^{(l)}(\omega_p) \cr A_{p_B,\alpha}^{(l)}(\omega_p) },
  \nonumber \\
  & & \alpha=TE,TM; \;\; l=1,\ldots,N.
\label{8}
\end{eqnarray}
We note that the coefficients $ A_{p_F,\alpha}^{(l)} $ and $
A_{p_B,\alpha}^{(l)} $ for $ l=1,\ldots,N $ describe the
corresponding electric-field amplitudes at the beginning of $ l
$th layer.

Assuming TE and TM waves, the boundary transfer matrices $ {\cal
T}_{p,TE}^{(l)} $ and $ {\cal T}_{p,TM}^{(l)} $ have the form:
\begin{eqnarray}  
 {\cal T}_{p,TE}^{(l)}(\omega_p) &=& \nonumber \\
 & & \hspace{-1.5cm} \frac{1}{2} \pmatrix{ 1+
  f^{(l)}_p(\omega_p) g^{(l)}_p(\omega_p)
  & 1- f^{(l)}_p(\omega_p) g^{(l)}_p(\omega_p)
  \cr 1-f^{(l)}_p(\omega_p) g^{(l)}_p(\omega_p)
  & 1+ f^{(l)}_p(\omega_p) g^{(l)}_p(\omega_p) }, \nonumber \\
 {\cal T}_{p,TM}^{(l)}(\omega_p) &=& \nonumber \\
  & & \hspace{-1.5cm} \frac{1}{2} \pmatrix{
  f^{(l)}_p(\omega_p) + g^{(l)}_p(\omega_p)
  & f^{(l)}_p(\omega_p) - g^{(l)}_p(\omega_p)
  \cr f^{(l)}_p(\omega_p) - g^{(l)}_p(\omega_p)
  & f^{(l)}_p(\omega_p) + g^{(l)}_p(\omega_p) },
   \nonumber \\
  & &  \;\; l=0,\ldots,N;
\label{9}
\end{eqnarray}
$ f^{(l)}_p = \cos(\vartheta_p^{(l)}) / \cos(\vartheta_p^{(l+1)})
 $ and $ g^{(l)}_p = n_p^{(l)}/n_p^{(l+1)} $. The free-field
propagation matrices $ {\cal P}_p^{(l)}(\omega_p) $ can be written
as:
\begin{eqnarray}  
 {\cal P}_p^{(l)}(\omega_p) &=& \pmatrix{ \exp(i{\bf k}_{p,z}^{(l)}L_l) & 0 \cr
 0 & \exp(-i{\bf k}_{p,z}^{(l)}L_l) }, \nonumber \\
 & &  l=1,\ldots,N.
\label{10}
\end{eqnarray}

The positive-frequency electric-field operators $ \hat{\bf
E}_s^{(+)}(z,t) $ and $ \hat{\bf E}_i^{(+)}(z,t) $ for the signal
and idler fields can be decomposed into TE- and TM-wave
contributions $ \hat{\bf E}_{s,TE}^{(+)}(z,t) $, $ \hat{\bf
E}_{s,TM}^{(+)}(z,t) $, $ \hat{\bf E}_{i,TE}^{(+)}(z,t) $, and $
\hat{\bf E}_{i,TM}^{(+)}(z,t) $ and expressed as follows
\cite{Vogel2001}:
\begin{eqnarray}    
 \hat{\bf E}_m^{(+)}(z,t) &=& \sum_{\alpha=TE,TM} \int_{0}^{\infty} d\omega_m \sqrt{
  \frac{\hbar\omega_m}{4\pi\epsilon_0 c{\cal B}} } \nonumber \\
  & & \mbox{} \hspace{0cm}\times \left[ \hat{a}_{m_F,\alpha}(z,\omega_m)
  {\bf e}_{m_F,\alpha}(z,\omega_m) \right. \nonumber \\
  & & \mbox{} \left. +
  \hat{a}_{m_B,\alpha}(z,\omega_m)
  {\bf e}_{m_B,\alpha}(z,\omega_m) \right]
  \exp(-i\omega_m t)  \nonumber \\
 & & \hspace{-17mm} =\hat{\bf E}_{m,TE}^{(+)}(z,t) + \hat{\bf
 E}_{m,TM}^{(+)}(z,t)
  \nonumber \\
 & & \hspace{-17mm} = \frac{1}{\sqrt{2\pi}} \int_{0}^{\infty}
  d\omega_m \hat{\bf E}_m^{(+)}(z,\omega_m) \nonumber \\
 & & \hspace{-17mm} = \frac{1}{\sqrt{2\pi}} \int_{0}^{\infty}
  d\omega_m \left[ \hat{\bf E}_{m,TE}^{(+)}(z,\omega_m) +
  \hat{\bf E}_{m,TM}^{(+)}(z,\omega_m) \right],
  \nonumber \\
 & & m=s,i .
 \label{11}
\end{eqnarray}
The permittivity of vacuum is denoted as $ \epsilon_0 $; $ c $
means speed of light in vacuum, $ \hbar $ is the reduced Planck
constant, and $ \cal B $ denotes the area of the transverse
profile of a beam. The symbols $ {\bf e}_{m_F,\alpha}(z,\omega_m)
$ and $ {\bf e}_{m_B,\alpha}(z,\omega_m) $ mean polarization
vectors of mode $ m $ of $ \alpha $ wave propagating forward and
backward with respect to the $ z $ axis. Annihilation operators of
signal [$ \hat{a}_{s_F,\alpha}(z,\omega_s) $, $
\hat{a}_{s_B,\alpha}(z,\omega_s) $] and idler [$
\hat{a}_{i_F,\alpha}(z,\omega_i) $, $
\hat{a}_{i_B,\alpha}(z,\omega_i) $] photons in $ \alpha $ wave
introduced in Eq. (\ref{11}) can be expressed in the same way as
the pump-field amplitude $ {\bf E}_{p,\alpha}^{(+)}(z,\omega_p) $:
\begin{eqnarray}     
 \hat{a}_{m_F,\alpha}(z,\omega_m){\bf e}_{m_F,\alpha}(z,\omega_m) +
  \hat{a}_{m_B,\alpha}(z,\omega_m){\bf e}_{m_B,\alpha}(z,\omega_m)
  && \nonumber \\
 & & \hspace{-8cm} = {\rm rect}_{-\infty,z_0}(z)
  \nonumber \\
 & &  \hspace{-8cm} \mbox{} \times  \left[
  \hat{a}_{m_F,\alpha}^{(0)}(\omega_m) {\bf e}_{m_F,\alpha}^{(0)}(\omega_m)
  \exp(i{\bf k}_{m,z}^{(0)}(z-z_0)) \right. \nonumber \\
 & & \hspace{-8cm} \mbox{}  \left. + \hat{a}_{m_B,\alpha}^{(0)}(\omega_m)
  {\bf e}_{m_B,\alpha}^{(0)}(\omega_m)\exp(-i{\bf k}_{m,z}^{(0)}(z-z_0)) \right]
  \nonumber \\
 & & \hspace{-8cm} \mbox{} + \sum_{l=1}^{N} {\rm rect}_{z_{l-1},z_l}(z) \nonumber \\
 & & \hspace{-8cm} \mbox{} \times \left[
  \hat{a}_{m_F,\alpha}^{(l)}(\omega_m) {\bf e}_{m_F,\alpha}^{(l)}(\omega_m)
  \exp(i{\bf k}_{m,z}^{(l)}(z-z_{l-1})) \right. \nonumber \\
 & & \hspace{-8cm} \mbox{}  \left. + \hat{a}_{m_B,\alpha}^{(l)}(\omega_m)
  {\bf e}_{m_B,\alpha}^{(l)}(\omega_m) \exp(-i{\bf k}_{m,z}^{(l)}(z-z_{l-1})) \right]
  \nonumber \\
 & & \hspace{-8cm}\mbox{} + {\rm rect}_{z_N,\infty}(z) \nonumber \\
 & & \hspace{-8cm} \mbox{} \times \left[
  \hat{a}_{m_F,\alpha}^{(N+1)}(\omega_m) {\bf e}_{m_F,\alpha}^{(N+1)}(\omega_m)
  \exp(i{\bf k}_{m,z}^{(N+1)}(z-z_N)) \right. \nonumber \\
 & & \hspace{-8cm}\mbox{} \left. + \hat{a}_{m_B,\alpha}^{(N+1)}(\omega_m)
  {\bf e}_{m_B,\alpha}^{(N+1)}(\omega_m) \exp(-i{\bf k}_{m,z}^{(N+1)}(z-z_N))
  \right],\nonumber \\
 & & \hspace{-6cm} m=s,i; \;\; \alpha=TE,TM.
\label{12}
\end{eqnarray}
The polarization vectors $ {\bf e}_{m_F,\alpha}^{(l)} $ and $ {\bf
e}_{m_B,\alpha}^{(l)} $ give the polarization directions of mode $
m $ with $ \alpha $ wave in $ l $th layer propagating forward and
backward, respectively, whereas $ {\bf k}_{m,z}^{(l)} $ is a $ z $
component of the wave vector $ {\bf k}_{m}^{(l)} $ of this mode:
\begin{equation}    
 {\bf k}_{m,z}^{(l)} = {\bf k}_{m}^{(l)} \cos(\vartheta_m^{(l)}),
 \hspace{5mm} m=s,i.
\end{equation}
The angle $ \vartheta_m^{(l)} $ characterizes the direction of
propagation of mode $ m $ with respect to the $ z $ axis. The angles $
\vartheta_m^{(l)} $ are given by the Snell law at the boundaries,
i.e.
\begin{eqnarray}  
 n_m^{(l)}\sin(\vartheta_m^{(l)}) &=&
 n_m^{(l+1)}\sin(\vartheta_m^{(l+1)}),
  \nonumber \\
  & &
  \hspace{-2cm} m=s,i; \;\; l=0,\ldots,N;
\end{eqnarray}
$ \vartheta_m^{(0)} = \vartheta_m $, where $ \vartheta_m $ stands
for the angle of incidence of mode $ m $.

The operators $ \hat{a}_{m_F,\alpha}^{(l)}(\omega_m) $ and $
\hat{a}_{m_B,\alpha}^{(l)}(\omega_m) $ obey the following
commutation relations:
\begin{eqnarray}   
 [ \hat{a}_{m_F,\alpha}^{(l)}(\omega_m),
   \hat{a}_{m'_F,\alpha'}^{(l')\dagger}(\omega'_m)]
   &=& \delta_{\alpha,\alpha'} \delta_{m,m'} \delta_{l,l'} \delta(\omega_m-\omega'_m) ,
   \nonumber \\
 {} [ \hat{a}_{m_F,\alpha}^{(l)}(\omega_m),
   \hat{a}_{m'_F,\alpha'}^{(l')}(\omega'_m)] &=& 0 ,
   \nonumber \\
 {} [ \hat{a}_{m_B,\alpha}^{(l)}(\omega_m),
   \hat{a}_{m'_B,\alpha'}^{(l')\dagger}(\omega'_m)]
   &=& \delta_{\alpha,\alpha'}\delta_{m,m'} \delta_{l,l'} \delta(\omega_m-\omega'_m) ,
   \nonumber \\
 {} [ \hat{a}_{m_B,\alpha}^{(l)}(\omega_m),
   \hat{a}_{m'_B,\alpha'}^{(l')}(\omega'_m)] &=& 0 ,
   \nonumber \\
 {} [ \hat{a}_{m_F,\alpha}^{(l)}(\omega_m),
   \hat{a}_{m'_B,\alpha'}^{(l')}(\omega'_m) ] &=& 0 ,
   \nonumber \\
 {} [ \hat{a}_{m_F,\alpha}^{(l)}(\omega_m),
   \hat{a}_{m'_B,\alpha'}^{(l')\dagger}(\omega'_m)] &=& 0 ;
   \nonumber \\
   & & \hspace{-4.5cm} m,m'=s,i; \;\; \alpha=TE,TM; \; \; l,l'=0,\ldots,N+1 .
\end{eqnarray}

The photonic-band-gap structure imposes the following relations
among the operators $ \hat{a}_{m_F,\alpha}^{(l)}(\omega_m) $ and $
\hat{a}_{m_B,\alpha}^{(l)}(\omega_m) $ acting in $ l $th layer:
\begin{eqnarray}  
 \pmatrix{ \hat{a}_{m_F,\alpha}^{(1)}(\omega_m) \cr \hat{a}_{m_B,\alpha}^{(1)}(\omega_m) } &=& {\cal
  T}_{m,\alpha}^{(0)}(\omega_m) \pmatrix{ \hat{a}_{m_F,\alpha}^{(0)}(\omega_m) \cr
  \hat{a}_{m_B,\alpha}^{(0)}(\omega_m)}, \nonumber \\
 \pmatrix{ \hat{a}_{m_F,\alpha}^{(l+1)}(\omega_m) \cr \hat{a}_{m_B,\alpha}^{(l+1)}(\omega_m) } &=&
  {\cal T}_{m,\alpha}^{(l)} {\cal P}_m^{(l)}  \pmatrix{ \hat{a}_{m_F,\alpha}^{(l)}(\omega_m)
  \cr \hat{a}_{m_B,\alpha}^{(l)}(\omega_m) }, \nonumber \\
  & & \hspace{-2cm}  m=s,i; \;\; \alpha=TE,TM; \;\; l=1,\ldots,N.
\end{eqnarray}
Transfer matrices $ {\cal T}^{(l)}_{m,\alpha} $ at the boundaries
and free-field propagation matrices $ {\cal P}^{(l)}_m $ for $
m=s,i $ and $ \alpha=TE,TM $ are defined in the same way as those
given in Eqs. (\ref{9}) and (\ref{10}) for the pump-field
amplitudes.

\subsection{Nonlinear interaction inside the photonic-band-gap
structure}

The Hamiltonian $ \hat{H}(t) $ describing spontaneous parametric
down-conversion in a nonlinear medium of volume $ \cal V $ at time
$ t $ can be written as:
\begin{eqnarray}  
 \hat{H}(t) &=& \epsilon_0  \int_{\cal V} d{\bf r} \;  \nonumber \\
 & & \hspace{-10mm}  {\bf d}({\bf r}):
 \left[ {\bf E}_{p}^{(+)}({\bf r},t) \hat{\bf E}_{s}^{(-)}({\bf
r},t)
 \hat{\bf E}_{i}^{(-)}({\bf r},t) + {\rm h.c.} \right],
\label{17}
\end{eqnarray}
where $ {\bf d} $ denotes a third-order tensor of nonlinear
coefficients, the symbol $ : $ is shorthand of the  tensor $ {\bf
d} $ with respect to its three indices, and $ {\rm h.c.} $ stands
for a hermitian conjugated term. The negative-frequency
electric-field operators $ \hat{\bf E}_{m}^{(-)} $ for $ m=s,i $
have been introduced in Eq.~(\ref{17}) ($ \hat{\bf E}_{m}^{(-)}=
\hat{\bf E}_{m}^{(+)\dagger} $). Decomposing the interacting
fields into TE and TM waves, and using the inverse Fourier
transformation in Eq.~(\ref{17}), we
arrive at:
\begin{eqnarray}  
 \hat{H}(t) &=& \frac{\epsilon_0 {\cal B}}{\sqrt{2\pi}} \int_{0}^{L} dz
  \int_{0}^{\infty} d\omega_p \int_{0}^{\infty} d\omega_s
  \int_{0}^{\infty} d\omega_i \nonumber \\
 & & \hspace{-2cm} \sum_{\alpha,\beta,\gamma=TE,TM}
  {\bf d}(z) :
  \Bigl[ {\bf E}_{p,\alpha}^{(+)}(z,\omega_p) \hat{\bf E}_{s,\beta}^{(-)}(z,\omega_s)
  \hat{\bf E}_{i,\gamma}^{(-)}(z,\omega_i) \nonumber \\
 & &  + {\rm h.c.} \Bigr].
\label{18}
\end{eqnarray}
Integrations over the variables $ x $ and $ y $ in Eq.~(\ref{17})
impose conditions for $ x $ and $ y $ component of wave-vectors; $
\delta({\bf k}_{s,x}+{\bf k}_{i,x}) $ [$ {\bf k}_{p,x} = 0 $ is
assumed] and $ \delta({\bf k}_{s,y}+{\bf k}_{i,y}-{\bf k}_{p,y})
$. The latter $ \delta $-function provides the following relation
between the angles $ \vartheta_s^{(l)} $ and $ \vartheta_i^{(l)} $
of modes $ s $ and $ i $ in the $ l $th layer:
\begin{eqnarray}   
 \vartheta_i^{(l)} &=& \arcsin \left[ \frac{\omega_p}{\omega_i}
  \sin(\vartheta_p^{(l)}) - \frac{\omega_s}{\omega_i}
  \sin(\vartheta_s^{(l)}) \right] ,
  \nonumber \\
 & & l=0,\ldots,N+1 .
\label{19}
\end{eqnarray}

Solution of the Schr\"{o}dinger equation to first order in
nonlinear perturbation together with the assumption of incident
vacuum state $ |{\rm vac}\rangle $ in signal and idler fields
provides the output state $ |\psi\rangle_{s,\beta,i,\gamma}^{\rm
out} $ of signal field with $ \beta $ polarization and idler field
with $ \gamma $ polarization in the form:
\begin{eqnarray}    
 |\psi\rangle_{s,\beta,i,\gamma}^{\rm out} &=& |{\rm vac}\rangle - \frac{i}{2\sqrt{2\pi}c}
  \sum_{l=1}^{N} \int_{0}^{\infty} \, d\omega_p \int_{0}^{\infty} \, d\omega_s \sqrt{\omega_s}
  \nonumber \\
 & & \hspace{-1cm} \mbox{} \times \int_{0}^{\infty} \, d\omega_i
  \sqrt{\omega_i} \sum_{l=1}^{N}
  \sum_{m=p_F,p_B} \sum_{n=s_F,s_B} \sum_{o=i_F,i_B} \nonumber \\
 & & \hspace{-1cm} \mbox{} \times \sum_{\alpha=TE,TM}
  {\bf d}^{(l)}:
  {\bf e}_{m,\alpha}^{(l)}(\omega_p) {\bf e}_{n,\beta}^{(l)}(\omega_s)
  {\bf e}_{o,\gamma}^{(l)}(\omega_i)  \nonumber \\
 & & \hspace{-1cm} \mbox{} \times A_{m,\alpha}^{(l)}(\omega_p)
  L_l\exp\left[+\frac{i}{2}(K_m^{(l)} - K_n^{(l)} - K_o^{(l)} )L_l \right]
  \nonumber \\
 & & \hspace{-1cm}\mbox{} \times
  {\rm sinc}\left[\frac{1}{2}(K_m^{(l)} - K_n^{(l)} - K_o^{(l)} )L_l \right]
  \delta(\omega_p-\omega_s-\omega_i)
  \nonumber \\
 & & \hspace{-1cm} \mbox{} \times
  \hat{a}_{n,\beta}^{(l)\dagger}(\omega_s) \hat{a}_{o,\gamma}^{(l)\dagger}(\omega_i)
  |{\rm vac} \rangle.
  \label{20}
\end{eqnarray}
The wave-vectors $ K^{(l)} $ introduced in Eq. (\ref{20}) are
defined as $ K_{j_F}^{(l)} = {\bf k}_{j,z}^{(l)} $ and $
K_{j_B}^{(l)} = - {\bf k}_{j,z}^{(l)} $ for $ j=p,s,i $.

The operators $ \hat{a}_{m_F,\alpha}^{(l)} $ and $
\hat{a}_{m_B,\alpha}^{(l)} $ for $ \alpha $ waves in mode $ m $ in
$ l $th layer can be expressed in terms of the operators $
\hat{a}_{m_F,\alpha}^{(N+1)} $ and $ \hat{a}_{m_B,\alpha}^{(0)} $.
These relations can be, e.g., written in the form:
\begin{eqnarray}   
 \pmatrix{  \hat{a}_{m_F,\alpha}^{(l)}(\omega_m) \cr  \hat{a}_{m_B,\alpha}^{(l)}(\omega_m)} &=&
  \prod_{j=l}^{1} \left[{\cal P}_m^{(j)}(\omega_m) {\cal T}_{m,\alpha}^{(j-1)}(\omega_m)
  \right] \nonumber \\
 & & \hspace{-3cm}\mbox{}  \times \pmatrix{ 1/({\cal S}_{m,\alpha})_{11}(\omega_m) &
  - ({\cal S}_{m,\alpha})_{12}(\omega_m) /
  ({\cal S}_{m,\alpha})_{11}(\omega_m) \cr 0 & 1 } \nonumber \\
 & & \hspace{-3cm} \mbox{} \times
  \pmatrix{\hat{a}_{m_F,\alpha}^{(N+1)}(\omega_m) \cr  \hat{a}_{m_B,\alpha}^{(0)}(\omega_m)} ,
  \nonumber \\
 & & \hspace{-2cm} m=s,i; \;\; \alpha=TE,TM; \;\; l=1,\ldots,N.
\label{21}
\end{eqnarray}
The matrix $ {\cal S}_{m,\alpha} $ used in Eq. (\ref{21})
describes the propagation of $ \alpha $ wave in field $ m $
through the whole structure:
\begin{eqnarray}    
 {\cal S}_{m,\alpha}(\omega_m) &=& {\cal T}_{m,\alpha}^{(N)}(\omega_m) \prod_{j=N}^{1}
  \left[{\cal P}_m^{(j)}(\omega_m)
  {\cal T}_{m,\alpha}^{(j-1)}(\omega_m) \right], \nonumber \\
  & &  m=s,i; \;\; \alpha=TE,TM.
 \label{22}
\end{eqnarray}

Similarly, the pump-field amplitudes $ A_{p_F,\alpha}^{(l)} $ and
$ A_{p_B,\alpha}^{(l)} $ for $ \alpha $ waves in $ l $th layer can
be determined from the amplitudes of the incident fields $
A_{p_F,\alpha}^{(0)} $ and $ A_{p_B,\alpha}^{(N+1)} $ as follows:
\begin{eqnarray}   
 \pmatrix{ A_{p_F,\alpha}^{(l)}(\omega_p) \cr A_{p_B,\alpha}^{(l)}(\omega_p)} &=&
 \prod_{j=l}^{1} \left[{\cal P}_p^{(j)}(\omega_p) {\cal T}_{p,\alpha}^{(j-1)}(\omega_p)
 \right] \nonumber \\
 & & \hspace{-3cm}\mbox{}  \times \pmatrix{ 1 & 0 \cr
 - ({\cal S}_{p,\alpha})_{21}(\omega_p) /({\cal S}_{p,\alpha})_{22}(\omega_p) &
 1/({\cal S}_{p,\alpha})_{22}(\omega_p) } \nonumber \\
 & & \hspace{-3cm} \mbox{} \times
  \pmatrix{A_{p_F,\alpha}^{(0)}(\omega_p) \cr A_{p_B,\alpha}^{(N+1)}(\omega_p)} ,
  \nonumber \\
 & & \hspace{-.5cm} \alpha = TE,TM; \;\;   l=1,\ldots,N .
\label{23}
\end{eqnarray}
The matrix $ {\cal S}_{p,\alpha} $ describes the propagation of a
classical pump field with polarization $ \alpha $ through the
whole structure, i.e.
\begin{equation}    
 {\cal S}_{p,\alpha}(\omega_p) = {\cal T}_{p,\alpha}^{(N)}(\omega_p) \prod_{j=N}^{1}
  \left[{\cal P}_p^{(j)}(\omega_p) {\cal T}_{p,\alpha}^{(j-1)}(\omega_p) \right].
\label{24}
\end{equation}

We note that the expression in Eq.~(\ref{20}) for the output state
$ |\psi\rangle_{s,\beta,i,\gamma}^{\rm out} $ including relations
written in Eqs. (\ref{21}-\ref{24}) can be formally recast into a
compact form using the so-called left-to-right ($ \Phi^{(+)} $)
and right-to-left ($ \Phi^{(-)} $) modes introduced in the
classical electro-magnetic theory of layered structures (for
details, see \cite{Centini2005}). Fields exiting the structure at
$ z=z_N $ are described by $ \Phi^{(+)}(z) $ functions, whereas $
\Phi^{(-)}(z) $ functions are appropriate for fields exiting at $
z=z_0 $. In classical theory, this corresponds to the picture in
which every nonlinear layer is a source (emitting 'dipole') of
photon pairs \cite{Aguanno2004} and the fields leaving the
structure are given as a sum of contributions from all layers.

We assume that the outgoing signal (idler) field is detected using
an analyzer with the polarization that forms an angle $
\varphi_s $ ($ \varphi_i $) with respect to the TE-wave polarization
direction (see Fig.~\ref{fig2}). In order to get the right
wave-function, we transform the operators of the outgoing signal
(idler) fields into the basis with the polarization vectors given
by angles $ \varphi_s $ ($ \varphi_i $) and $ \varphi_s +\pi/2 $
($ \varphi_i + \pi/2 $) using the following formulas:
\begin{eqnarray}   
 \pmatrix{ \hat{a}_{m_F,TE}^{(N+1)} \cr \hat{a}_{m_F,TM}^{(N+1)} } &=&
  \nonumber \\
  & & \hspace{-2cm}
  \pmatrix{ \cos(\varphi_m) & \sin(\varphi_m)
    \cr -\sin(\varphi_m) & \cos(\varphi_m) }
  \pmatrix{ \hat{a}_{m_F,\varphi_m}^{(N+1)} \cr
  \hat{a}_{m_F,\varphi_m+\pi/2}^{(N+1)} },
  \nonumber \\
 \pmatrix{ \hat{a}_{m_B,TE}^{(0)} \cr \hat{a}_{m_B,TM}^{(0)}} &=&
  \nonumber \\
  & & \hspace{-2cm}
  \pmatrix{ \cos(\varphi_m) & \sin(\varphi_m)
    \cr -\sin(\varphi_m) & \cos(\varphi_m)}
  \pmatrix{ \hat{a}_{m_B,\varphi_m}^{(0)} \cr
  \hat{a}_{m_B,\varphi_m+\pi/2}^{(0)} },
  \nonumber \\
 & & m=s,i .
\label{25}
\end{eqnarray}

Substituting relations in Eqs. (\ref{21}), (\ref{23}), and
(\ref{25}) into the expression $ |\psi\rangle_{s,TE,i,TE}^{\rm
out} + |\psi\rangle_{s,TE,i,TM}^{\rm out} +
|\psi\rangle_{s,TM,i,TE}^{\rm out} + |\psi\rangle_{s,TM,i,TM}^{\rm
out} $ determined using Eq.~(\ref{20}), we arrive at the output
state $ |\psi\rangle_{s,\varphi_s,i,\varphi_i}^{\rm out} $
describing a signal photon polarized along the angle $ \varphi_s $
and an idler photon polarized along the angle $ \varphi_i $:
\begin{eqnarray}    
 |\psi\rangle_{s,\varphi_s,i,\varphi_i}^{\rm out} &=&
  |{\rm vac}\rangle + \int \, d\omega_s  \int \, d\omega_i
  \biggl[ \nonumber \\
  & & \phi^{FF}(\omega_s,\omega_i)
  \hat{a}_{s_F,\varphi_s}^{(N+1)\dagger}
  \hat{a}_{i_F,\varphi_i}^{(N+1)\dagger} |{\rm vac} \rangle
   \nonumber \\
  & & \mbox{} + \phi^{FB}(\omega_s,\omega_i)
  \hat{a}_{s_F,\varphi_s}^{(N+1)\dagger}
  \hat{a}_{i_B,\varphi_i}^{(0)\dagger} |{\rm vac} \rangle
   \nonumber \\
   & & \mbox{} +
  \phi^{BF}(\omega_s,\omega_i)
  \hat{a}_{s_B,\varphi_s}^{(0)\dagger}
  \hat{a}_{i_F,\varphi_i}^{(N+1)\dagger} |{\rm vac} \rangle
   \nonumber \\
   & & \mbox{} +
  \phi^{BB}(\omega_s,\omega_i)
  \hat{a}_{s_B,\varphi_s}^{(0)\dagger}
  \hat{a}_{i_B,\varphi_i}^{(0)\dagger} |{\rm vac} \rangle
  \biggr] .
\label{26}
\end{eqnarray}
The function $ \phi^{mn}(\omega_s,\omega_i) $ introduced in Eq.
(\ref{26}) has the meaning of probability amplitude of having a
signal photon in field $ m $ polarized along the angle $ \varphi_s
$ and its entangled idler twin in field $ n $ polarized along the
angle $ \varphi_i $ at the output from the photonic-band-gap
structure.

We are interested only in the part $ |\psi\rangle_{s,i}^{(2)} $ of
the output state $ |\psi\rangle_{s,\varphi_s,i,\varphi_i}^{\rm
out} $ given in Eq. (\ref{26}) that describes the generated
photon-pairs. Including the time evolution of the free-fields
outside the photonic-band-gap structure, we can write:
\begin{equation}    
 |\psi(t)\rangle_{s,i}^{(2)} = |\psi_{s,i}^{FF}(t) \rangle +
 |\psi_{s,i}^{FB}(t) \rangle + |\psi_{s,i}^{BF}(t) \rangle +
 |\psi_{s,i}^{BB}(t) \rangle,
\end{equation}
where
\begin{eqnarray}   
 |\psi_{s,i}^{mn}(t)\rangle &=&
  \int_{0}^{\infty} \, d\omega_s  \int_{0}^{\infty} \, d\omega_i \;
  \phi^{mn}(\omega_s,\omega_i)
  \nonumber \\
 & & \mbox{} \hspace{-1.5cm}\times
  \hat{a}_{s_m}^{\dagger}(\omega_s)
  \hat{a}_{i_n}^{\dagger}(\omega_i) \exp(i\omega_s t) \exp(i
  \omega_i t)
  |{\rm vac} \rangle , \nonumber \\
 & & \hspace{1cm} m,n = F,B;
 \label{28}
\end{eqnarray}
The creation operators $ \hat{a}_{m}^\dagger $, $
m=s_F,i_F,s_B,i_B $, introduced in Eq.~(\ref{28}) describe the
linearly polarized photons outside the photonic-band-gap structure
with  polarization angles $ \varphi_s $ and $ \varphi_i $, and are
given as:
\begin{eqnarray}    
 \hat{a}_{s_F}^\dagger(\omega_s) =
 \hat{a}_{s_F,\varphi_s}^{(N+1)\dagger}(\omega_s), & &
 \hat{a}_{i_F}^\dagger(\omega_i)=
 \hat{a}_{i_F,\varphi_i}^{(N+1)\dagger}(\omega_i), \nonumber \\
 \hat{a}_{s_B}^\dagger(\omega_s) =
 \hat{a}_{s_B,\varphi_s}^{(0)\dagger}(\omega_s), & &
 \hat{a}_{i_B}^\dagger(\omega_i) =
 \hat{a}_{i_B,\varphi_i}^{(0)\dagger}(\omega_i).
\end{eqnarray}
We note that we do not explicitly express the dependence of
quantities on the polarization angles $ \varphi_s $ and $ \varphi_i $
below.

\subsection{Properties of the signal and idler fields outside the
structure}

The mean number $ N_{s,i}^{mn} $ of photon pairs that have a
signal photon in the frequency interval $ \Delta\omega_s $ around
frequency $ \omega_s $ and its twin idler photon in frequency
interval $ \Delta\omega_i $ around frequency $ \omega_i $ in field
$ mn $ is given by
\begin{equation}       
 N_{s,i}^{mn}(\omega_s,\omega_i) = \langle \psi_{s,i}^{mn}(t)| \hat{n}_{s_m}(\omega_s)
  \hat{n}_{i_n}(\omega_i)  |\psi_{s,i}^{mn}(t)\rangle \Delta \omega_s
  \Delta \omega_i ,
\label{30}
\end{equation}
where the operator $ \hat{n}_{j_m}(\omega_j) $, the density of
photons, is defined as
\begin{equation}       
 \hat{n}_{j_m}(\omega_j) = \hat{a}_{j_m}^\dagger(\omega_j)
  \hat{a}_{j_m}(\omega_j), \hspace{5mm} j=s,i.
\end{equation}
Using Eq. (\ref{28}), the expression for $ N_{s,i}^{mn} $ in Eq.
(\ref{30}) can be written as:
\begin{equation}    
 N_{s,i}^{mn}(\omega_s,\omega_i) =
 |\phi^{mn}(\omega_s,\omega_i)|^2 \Delta \omega_s \Delta \omega_i
 .
\end{equation}

We get the following expressions for, e.g., the mean number $
N_{s}^{mn}(\omega_s)$ of signal photons in frequency interval $
\Delta\omega_s $ around frequency $ \omega_s $:
\begin{eqnarray}   
 N_{s}^{mn}(\omega_s) &=& \langle \psi_{s,i}^{mn}(t)| \hat{n}_{s_m}(\omega_s)
 |\psi_{s,i}^{mn}(t)\rangle \Delta \omega_s  \nonumber \\
 &=& \int_{0}^{\infty} d\omega_i |\phi^{mn}(\omega_s,\omega_i)|^2 \Delta \omega_s.
\end{eqnarray}

The overall number $ N^{mn} $ of photon pairs emitted into field $
mn $ is determined by the expression
\begin{equation}    
 N^{mn} = \int_{0}^{\infty} d\omega_s \int_{0}^{\infty}
d\omega_i
  |\phi^{mn}(\omega_s,\omega_i)|^2.
\end{equation}

The signal-field energy spectrum $ S^{mn}_s(\omega_s) $ of field $
mn $ can be easily determined using the expression
\begin{eqnarray}         
 S^{mn}_s(\omega_s)  &=& \hbar \omega_s
  \frac{N_{s}^{mn}(\omega_s)}{\Delta \omega_s} \nonumber \\
 &=& \hbar \omega_s \int_{0}^{\infty} d\omega_i
 |\phi^{mn}(\omega_s,\omega_i)|^2.
\label{35}
\end{eqnarray}

The energy spectra $ S_{j_m}(\omega_j) $ ($ j=s,i $; $ m=F,B $)
characterizing the outgoing down-converted fields without an
inclusion of pair entanglement can be evaluated using the
signal-field spectra $ S^{mn}_{s} $ determined in Eq. (\ref{35});
the idler-field spectra $ S^{mn}_{i} $ are determined analogously;
\begin{eqnarray}   
 S_{s_F} &=& S^{FF}_{s_F} +
 S^{FB}_{s_F} , \nonumber \\
 S_{i_F} &=& S^{FF}_{i_F} +
 S^{BF}_{i_F} , \nonumber \\
 S_{s_B} &=& S^{BF}_{s_B} +
 S^{BB}_{s_B} , \nonumber \\
 S_{i_B} &=& S^{FB}_{i_B} +
 S^{BB}_{i_B} .
\end{eqnarray}

The properties of the down-converted fields in the time domain can be
conveniently described using a two-photon amplitude $ {\cal
A}(\tau_s,\tau_i) $ giving the probability amplitude of detecting
a signal photon at time $ \tau_s $ and an idler photon at time $
\tau_i $:
\begin{eqnarray}    
  {\cal A}^{mn}(\tau_s,\tau_i)
   &=& \langle {\rm vac} |
  \hat{E}^{(+)}_{s_m}(0,t_0+\tau_s) \nonumber \\
 & & \mbox{} \times \hat{E}^{(+)}_{i_n}(0,t_0+\tau_i)
  |\psi^{mn}_{s,i}(t_0)\rangle .
\label{37}
\end{eqnarray}
Assuming the state $ | \psi^{mn}_{s,i} \rangle $ given in Eq.
(\ref{28}), the expression in Eq. (\ref{37}) for the two-photon
amplitude $ {\cal A}^{mn} $ can be rearranged into the form:
\begin{equation}   
 {\cal A}^{mn}(\tau_s,\tau_i) = \frac{\hbar \sqrt{\omega_s^0 \omega_i^0}}{4\pi\epsilon_0 c {\cal B}}
 \phi^{mn}(\tau_s,\tau_i) ,
\end{equation}
where the `Fourier transform' $ \phi^{mn}(\tau_s,\tau_i) $ of the
function $ \phi^{mn}(\omega_s,\omega_i) $ has been introduced:
\begin{eqnarray}    
 \phi^{mn}(\tau_s,\tau_i) &=& \frac{1}{2\pi}
  \int_{0}^{\infty} d\omega_s \; \int_{0}^{\infty} d\omega_i \;
  \sqrt{\frac{\omega_s\omega_i}{\omega_s^0\omega_i^0}}
  \phi^{mn}(\omega_s,\omega_i)  \nonumber \\
  & & \mbox{} \times
 \exp(-i\omega_s\tau_s)
 \exp(-i\omega_i\tau_i).
\end{eqnarray}

The photon flux of, e.g., the signal photons, $ {\cal
N}^{mn}_{s}(\tau_s) $ at time $ \tau_s $ over the transverse
profile of area $ \cal B $ in field $ mn $ is defined as
\begin{eqnarray}       
 {\cal N}^{mn}_{s} (\tau_s) &=& \frac{\epsilon_0 c {\cal B}}{2} \langle \psi^{mn}_{s,i}(t_0) |
 \hat{E}_{s_m}^{(-)}(0,t_0 + \tau_s) \nonumber \\
 & & \mbox{} \times  \hat{E}_{s_m}^{(+)}(0,t_0 + \tau_s)
 | \psi^{mn}_{s,i}(t_0) \rangle.
\end{eqnarray}
The photon flux $ {\cal N}^{mn}_{s}(\tau_s) $ can be determined
using the function $ \phi^{mn}(\omega_s,\omega_i) $;
\begin{eqnarray}     
 {\cal N}^{mn}_{s}(\tau_s) &=& \frac{\hbar}{8\pi}  \int_{0}^{\infty}
  d\omega_s \; \sqrt{\omega_s} \int_{0}^{\infty}  d\omega_s' \; \sqrt{\omega_s'}
  \int_{0}^{\infty}  d\omega_i \;  \nonumber \\
  & & \hspace{-2.2cm} \mbox{} \times \phi^{mn*}(\omega_s',\omega_i)
   \phi^{mn}(\omega_s,\omega_i)
  \exp(i\omega_s'\tau_s) \exp(-i\omega_s\tau_s).
\end{eqnarray}
If the spectrum of the idler field is narrow, we may use the alternative
expression
\begin{equation}    
 {\cal N}^{mn}_{s}(\tau_s) = \frac{\hbar\omega_s^0}{4}
 \int_{-\infty}^{\infty} d\tau_i \; |\phi^{mn}(\tau_s,\tau_i)|^2 .
\end{equation}

\begin{figure}    
 \resizebox{0.8\hsize}{!}{\includegraphics{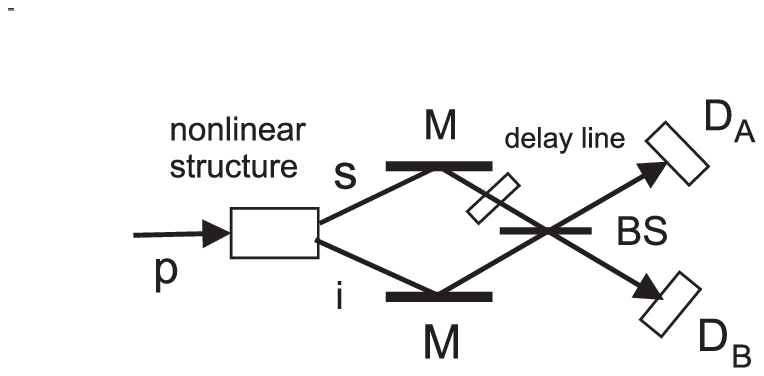}}
 \vspace{3mm}
 \caption{Scheme of Hong-Ou-Mandel interferometer.
 In a nonlinear structure a photon from a pump beam
 p is converted into a photon pair
 copropagating as signal (s) and idler (i) beams. After reflection
 from mirrors M and introduction of a mutual delay in a delay line, both photons
 interfere at a beam-splitter BS, and are then detected using  a coincidence-count
 detection scheme at detectors D${}_{\rm A}$ and D${}_{\rm B}$.}
\label{fig3}
\end{figure}

Entanglement of the signal and idler photons in a pair can be
detected in a Hong-Ou-Mandel interferometer (see Fig.~\ref{fig3}).
In order to achieve this type of interference between the signal and idler
photons, we rotate polarizations of both photons such that they
are the same, and subsequently introduce a relative time delay $ \tau_l $
between the two photons. Then two photons arrive at a 50/50~\% beam-splitter
whose output ports are monitored by detectors. The measured
coincidence-count rate $ R_c $ is given by the number of
simultaneously detected photons at detectors $ D_{\rm A} $ and $
D_{\rm B} $, placed at  the output ports of the beam-splitter in a given
time interval. There occurs quantum interference between two paths
leading to a coincidence count; either a signal photon is detected
by detector $ {\rm D}_A $ and its twin idler photon by detector $
{\rm D}_B $ or vice versa. The normalized coincidence-count rate $
R_n $ in this interferometer can be expressed as follows:
\begin{equation}    
 R_n^{mn}(\tau_l) = 1 - \rho^{mn}(\tau_l) ,
\end{equation}
where
\begin{eqnarray}   
 \rho^{mn}(\tau_l) &=& \frac{1}{2R_0^{mn}} \int_{-\infty}^{\infty} dt_A \,
 \int_{-\infty}^{\infty} dt_B \, \nonumber \\
 & & \hspace{-7mm} {\rm Re} \left[  {\cal A}^{mn}(t_A,
 t_B- \tau_l){\cal A}^{mn*}(t_B,t_A-\tau_l) \right]
\label{44}
\end{eqnarray}
and
\begin{equation}       
 R_0^{mn} = \frac{1}{2} \int_{-\infty}^{\infty} dt_A
 \int_{-\infty}^{\infty} dt_B
 \left| {\cal A}^{mn}(t_A,t_B) \right|^2.
\label{45}
\end{equation}
The symbol $ {\rm Re} $ stands for real part, and the two-photon
amplitude $ {\cal A}^{mn} $ has been introduced in Eq.~(\ref{37}).
Using the expressions in Eqs. (\ref{44}) and (\ref{45}),
we arrive at:
\begin{eqnarray}   
 \rho^{mn}(\tau_l) &=& \frac{\hbar^2}{8\epsilon_0^2 c^2 {\cal B}^2}
 \frac{1}{ R_0^{mn}}
 {\rm Re} \Biggl[ \int_{0}^{\infty} d\omega_s \, \int_{0}^{\infty} d\omega_i
 \; \omega_s \omega_i
 \nonumber \\
 & & \hspace{-12mm} \phi^{mn}(\omega_s,\omega_i)
  \phi^{mn*}(\omega_i,\omega_s)
 \exp(i\omega_i\tau_l) \exp(-i\omega_s\tau_l) \Biggr],
  \nonumber \\
  & &  \\
 R_0^{mn} &=& \frac{\hbar^2}{8\epsilon_0^2 c^2
  {\cal B}^2} \int_{0}^{\infty} d\omega_s \, \int_{0}^{\infty} d\omega_i \,
  \; \omega_s \omega_i
 |\phi^{mn}(\omega_s,\omega_i)|^2 . \nonumber \\
 & &
\end{eqnarray}

\subsection{Cw-limit}

If cw pumping is considered, the  spectrum is described using Eq.
(\ref{4}).  As a consequence the following expressions are
obtained:
\begin{equation}   
 |\phi^{mn}(\omega_s,\omega_i)|^2 =
 f(\omega_s,\omega_i)\delta^2(\omega_p-\omega_s-\omega_i).
\end{equation}
In this case the second power of the Dirac delta function $ \delta
$ is formal, so that the above expression should be replaced by:
\begin{equation}   
 |\phi^{mn}(\omega_s,\omega_i)|^2 =
 \lim_{T\rightarrow\infty} \frac{2T}{2\pi}
 f(\omega_s,\omega_i)\delta(\omega_p-\omega_s-\omega_i),
\end{equation}
where the period of nonlinear interaction goes from $ -T $ to $ T
$. Expressions for the physical quantities determined above have
to be normalized by $ 2T $, which indicates that these quantities are
related to the period of $ 1s $.

\section{Typical characteristics of the down-converted fields
generated from a photonic-band-gap structure}

As an example, we study the properties of a nonlinear
photonic-band-gap structure composed of 25 nonlinear layers of GaN
of thickness of 117~nm among which there are 24 linear layers of
AlN with a thickness of 180~nm. Material characteristics of GaN
and AlN can be found, e.g., in \cite{Sanford2005,Miragliotta1993}.
This structure is resonant for the pump field at wavelength of
664.5~nm, designed to correspond to the first resonance peak near
the band edge. This situation favors the efficient generation of
photon pairs at roughly double the pump wavelength, at angles that
correspond to the first, second, and third resonance peaks,
respectively, for the down-converted fields. We assume that the
GaN crystallographic $ x $ axis coincides with the $ z $ axis of
propagation (see Fig.~\ref{fig1}). In this configuration there is
strong nonlinear interaction between TE components of the pump,
signal, and idler beams. The nonlinear coefficient $ d $ for this
geometry is chosen to be 10 pmV$ \mbox{}^{-1} $.

\subsection{Cw pumping}

\begin{figure}    
 {\raisebox{4 cm}{a)} \hspace{0mm}
 \resizebox{0.9\hsize}{!}{\includegraphics{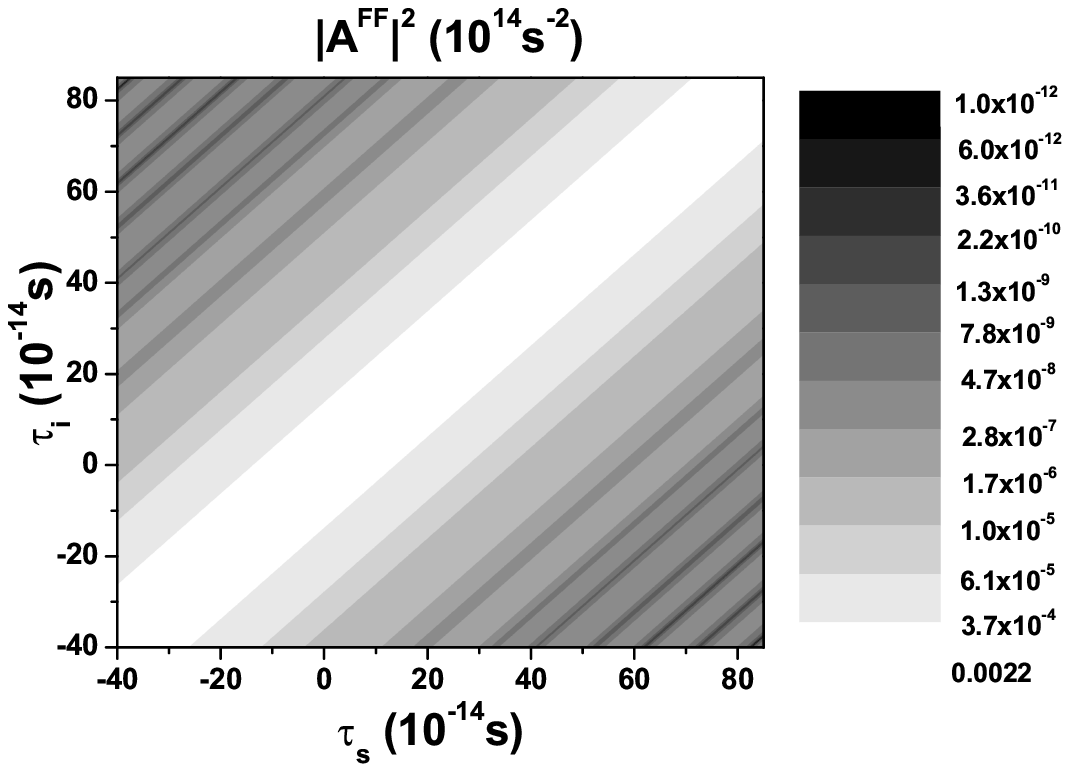}}

 \vspace{5mm}
 \hspace{-20mm} \raisebox{2 cm}{b)} \hspace{5mm}
 \resizebox{0.6\hsize}{0.3\hsize}{\includegraphics{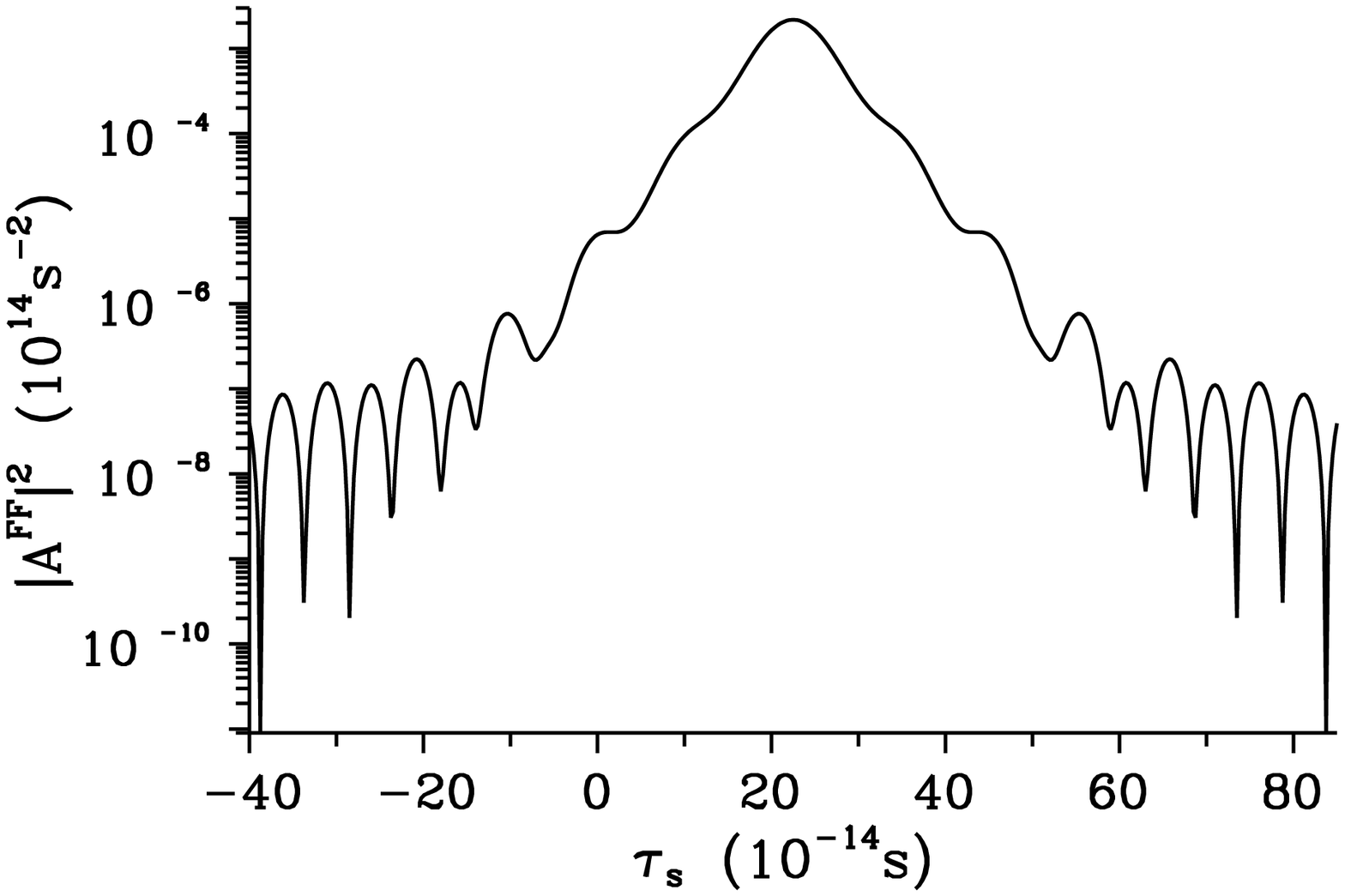}}}

 \vspace{10mm}

 \caption{Probability $ |{\cal A}^{FF}|^2 $ of
 detecting a signal photon at time $ \tau_s $ and its twin idler
 photon at time $ \tau_i $ (a) in mode $ FF $ for cw pumping is shown;
 cut of the graph along the line $ \tau_s + \tau_i = 450 $~fs is shown in (b);
 the probability $ |{\cal A}^{FF}|^2 $
 is normalized such that one photon pair is emitted, i.e.
 $ \int d\tau_s \int d\tau_i |{\cal A}^{FF}(\tau_s,\tau_i)|^2 = 1
 $: $ \theta_s = 14 $~deg; logarithmic scale on $ z $ axis is used.}
\label{fig4}
\end{figure}

We first consider pumping by a cw laser tuned at first resonance peak,
i.e. $ \lambda_p = 664.5 $~nm. A
photon pair whose signal photon is emitted along the angle $
\theta_s $ may be described by a two-photon amplitude $ \ {\cal A}
$ with typical shapes both in time and spectral domains. In the
spectral domain, the two-photon amplitude $ {\cal
A}(\omega_s,\omega_i) $ can be written in the form  $ s(\omega_s)
\delta(\omega_p^0 - \omega_s-\omega_i) $, where the function $
s(\omega_s) $ is linearly proportional to the spectral amplitude of the
signal field. This is a consequence of stationarity of the
emitted down-converted fields. The probability $ |{\cal
A}(\tau_s,\tau_i)|^2 $ of detecting a signal photon at time $
\tau_s $ and an idler photon at time $ \tau_i $ is shown in
Fig.~\ref{fig4} for mode $ FF $ (both photons exit the structure
at $ z=z_N $).

The shape of the probability $ |{\cal A}(\tau_s,\tau_i)|^2 $ can
be easily understood when we consider the fact that both photons
in a pair are born together at the same instant of time, and then
propagate independently inside the structure. This independent
zig-zag propagation increases the  average time delay between
the two photons, which takes into account all possible
realizations of the random process of propagation of a photon that
undergoes multiple bounces inside the structure. The 'random' zig-zag
propagation causes interference between different paths.  This
leads to a typical structure with local minima and maxima along
the direction $ \tau_s+\tau_i={\rm const} $; as shown in
Fig.~\ref{fig4}b. Also, the greater the difference $ \tau_s-\tau_i
$ the smaller the value of the two-photon amplitude $ {\cal A} $
because the propagation with more zig-zags is less probable. We
note that typical stripes along the direction $ \tau_s - \tau_i
={\rm const} $ reflect stationarity of the process.

\begin{figure}    
 {\raisebox{4 cm}{a)} \hspace{0mm}
 \resizebox{0.9\hsize}{!}{\includegraphics{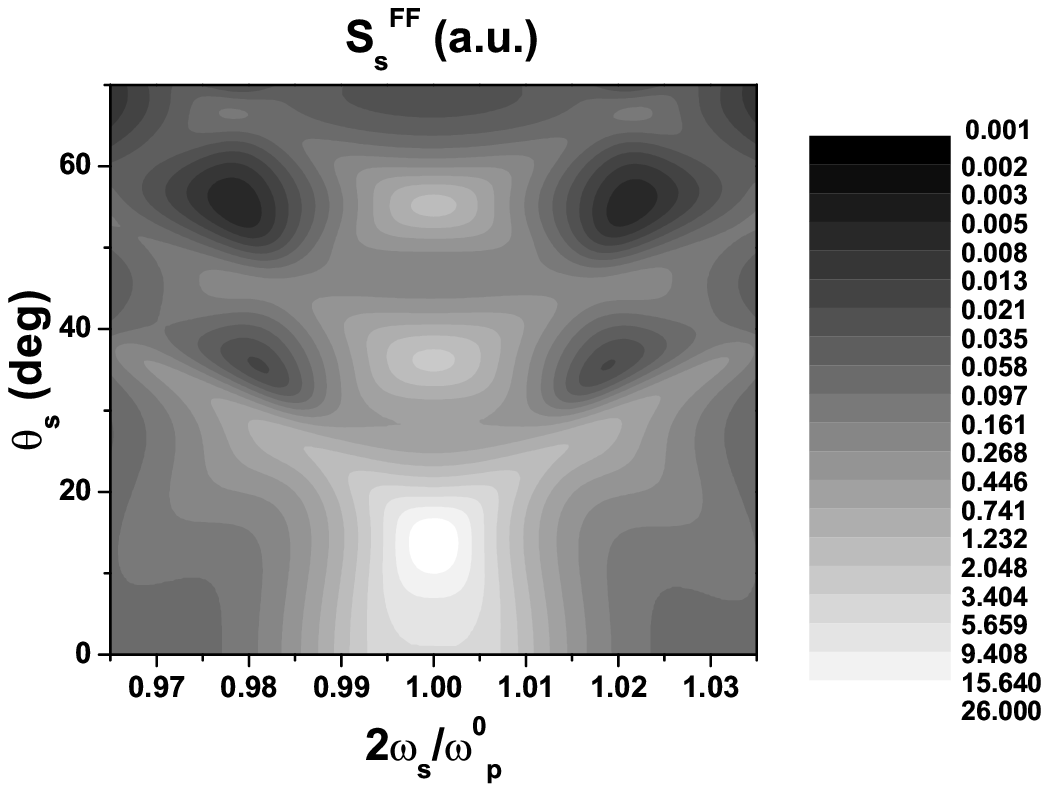}}

 \vspace{5mm}
 \hspace{-20mm} \raisebox{2 cm}{b)} \hspace{5mm}
 \resizebox{0.6\hsize}{0.3\hsize}{\includegraphics{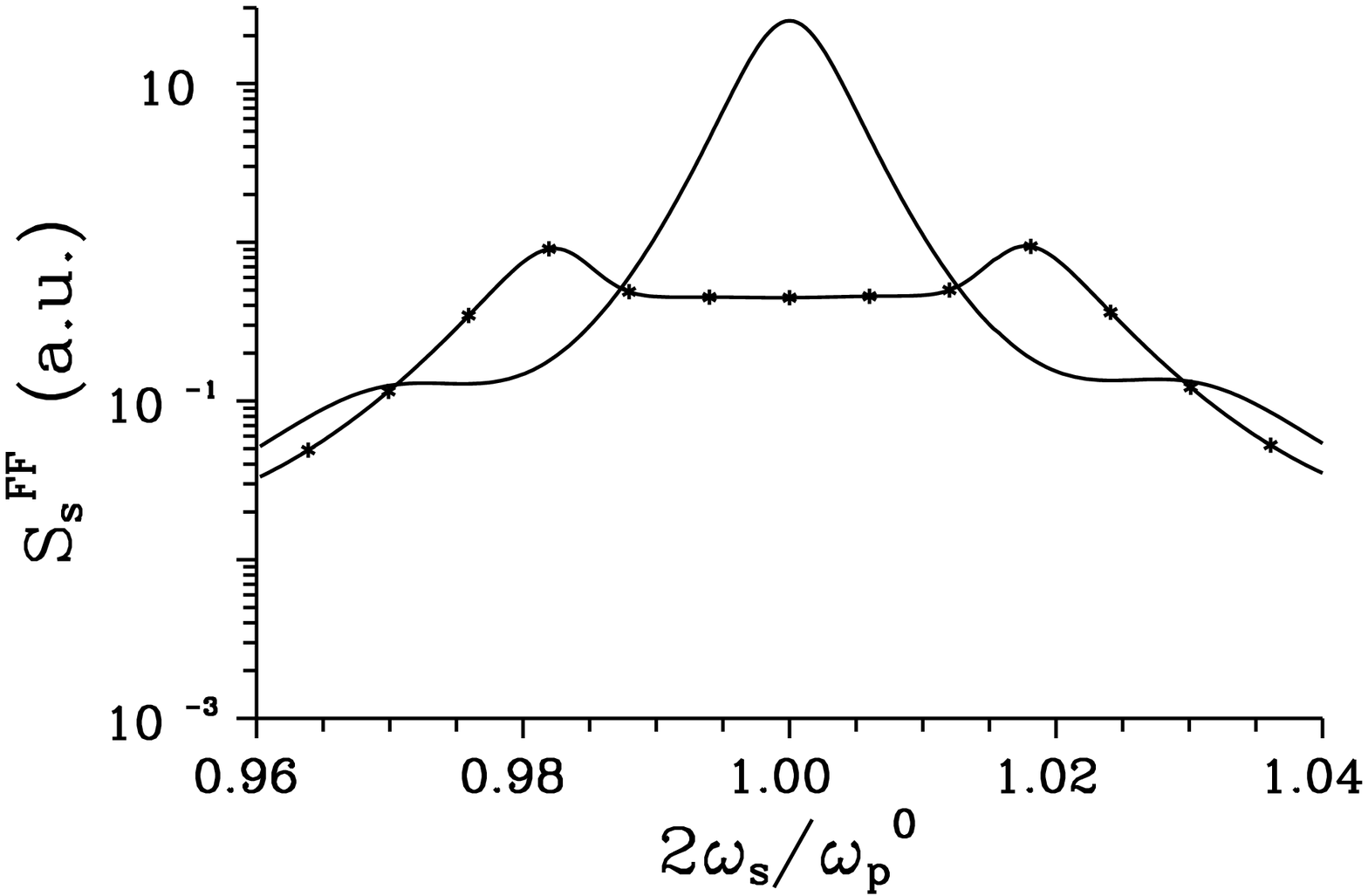}}}
 \vspace{2mm}

 \caption{Energy spectrum  $ S_s^{FF} $ of mode $ FF $ versus
 normalized signal-field frequency $ 2\omega_s/\omega_p^0 $ and angle
 $ \theta_s $ of emission of a signal photon is shown in (a); cuts along
 the lines $ \theta_s = 14 $~deg (solid curve) and $ \theta_s = 28 $~deg
 (solid curve with $ * $) are shown in (b); cw pumping is assumed;
 logarithmic scale is used on the $ z $ axis.}
\label{fig5}
\end{figure}

\begin{figure}    
 \resizebox{0.9\hsize}{!}{\includegraphics{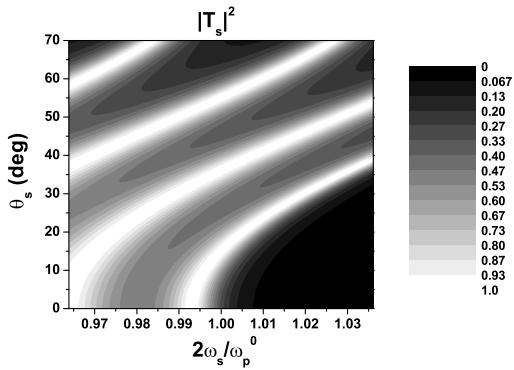}}
 \vspace{2mm}

 \caption{Signal-field intensity transmission $ |T_s|^2 $ as a
 function of normalized frequency $ 2\omega_s/\omega_p^0 $ and angle
 $ \theta_s $. Positions of peaks of resonance as they depend on
 the angle $ \theta_s $ are clearly visible.}
\label{fig6}
\end{figure}

Typical spectral properties of this structure are documented in
Fig.~\ref{fig5}, where the signal-field energy spectrum $ S_s^{FF}
$ is depicted as a function of the angle $ \theta_s $ for field $ FF $.
The spectrum $ S_s^{FF} $ giving the probability of emission of a
signal photon at frequency $ \omega_s $ along the angle $ \theta_s
$ is a complex function in its variables because it is built up by
interference of fields coming from the multilayer structure. A
maximum photon-pair generation rate is obtained for $ \theta_s
\approx 13.8 $~deg for degenerate frequencies ($ \omega_s \approx
\omega_i $). Photons from a generated photon pair can exit the
structure also at $ z=z_0 $; their energy spectra $
S_s^{FB}(\omega_s,\theta_s) $, $ S_s^{BF}(\omega_s,\theta_s) $ and
$ S_s^{BB}(\omega_s,\theta_s) $ are similar to that shown in
Fig.~\ref{fig5}. This is because efficient generation of photon
pairs occurs provided both the signal and idler fields are
resonant in the structure so that the generated photons exit the
structure at $ z=z_0 $ or $ z=z_N $ with comparable probabilities.
The signal-field intensity transmission $ |T_s|^2 $ as a function
of normalized frequency $ 2\omega_s/\omega_p^0 $ and angle $
\theta_s $ is shown in Fig.~\ref{fig6} for comparison.

\begin{figure}    
 {\raisebox{4 cm}{a)} \hspace{5mm}
 \resizebox{0.7\hsize}{!}{\includegraphics{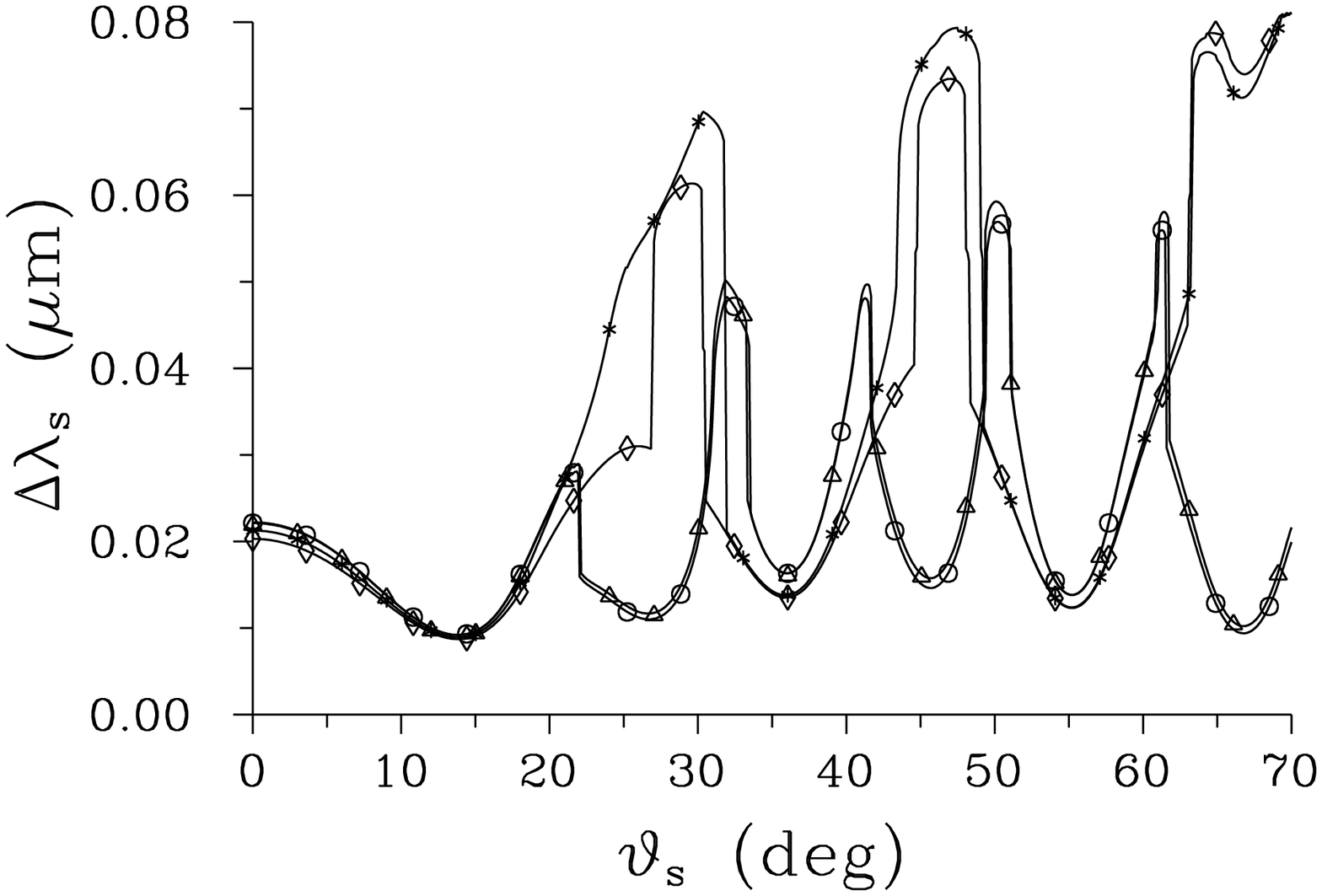}}

 \vspace{5mm}
 \raisebox{4 cm}{b)} \hspace{5mm}
 \resizebox{0.7\hsize}{!}{\includegraphics{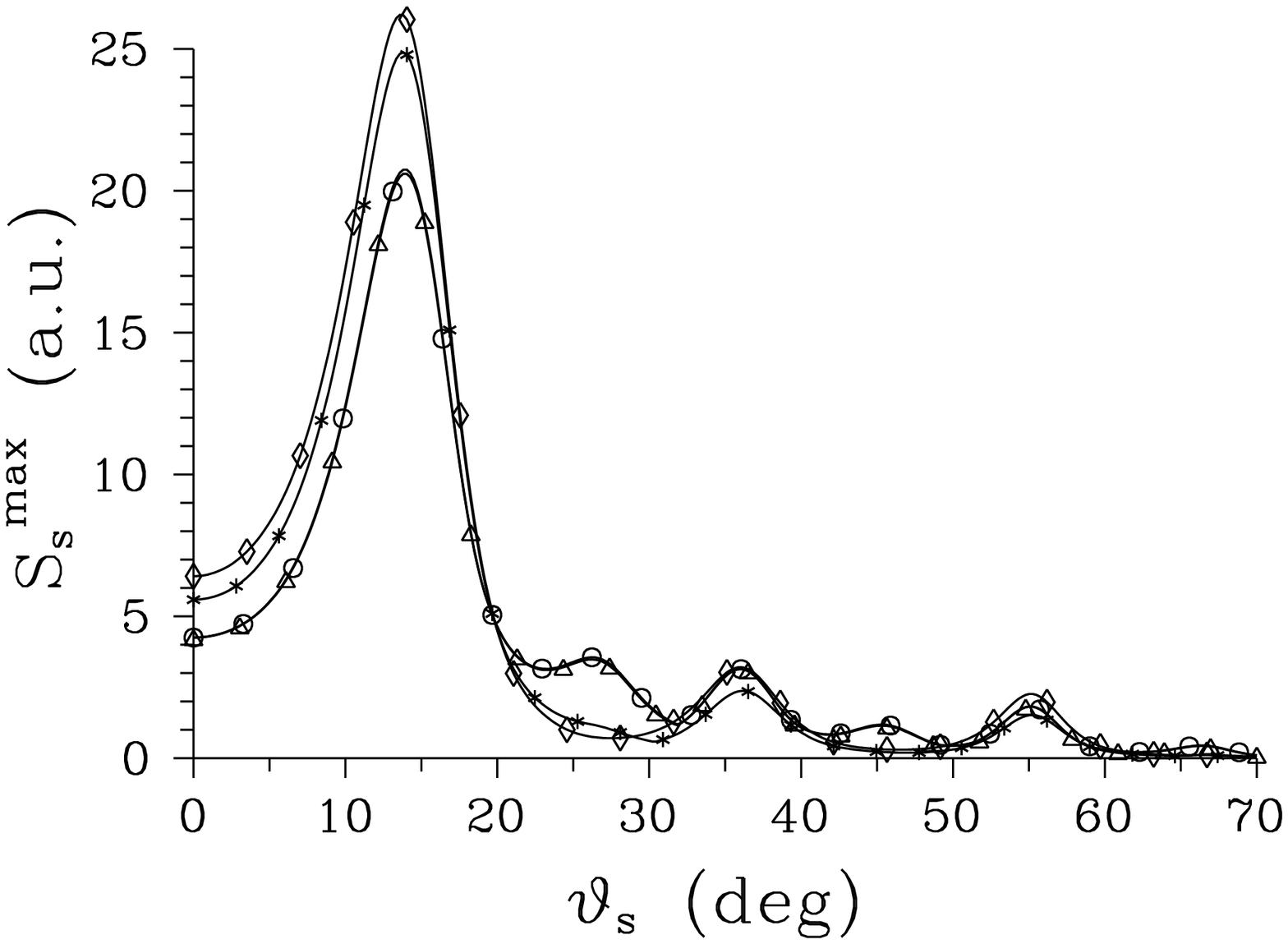}}}
 \vspace{2mm}

 \caption{FWHM $ \Delta\lambda_s $ (a) and maximum value $ S_s^{\rm max} $ at
 the central frequency $ \omega_s^c $ (b) for the signal-field energy spectrum
 $ S_s(\omega_s) $ of modes $ FF $ ($ * $), $ FB $ ($ \triangle $), $ BF $ ($ \circ $),
 and $ BB $ ($ \diamond $) as functions of the angle $ \theta_s $
 of emission of a signal photon are shown; cw pumping is assumed.}
\label{fig7}
\end{figure}

The angle-dependence of the signal-field FWHM (full-width at half
maximum) $ \Delta\lambda_s $ of the energy spectrum in wavelengths
and value of energy spectrum $ S_s^{\rm max} $ at the central
frequency $ \omega_s^c $ are shown in Fig.~\ref{fig7}. High
photon-pair generation rates are observed for the
frequency-degenerate case ($ \omega_s^c \approx \omega_i^c $) for
values of the angle $ \theta_s $ approximately equal to 14, 35,
and 55 degrees. For these values of the angle $ \theta_s $ the
peaks of the signal-field intensity transmission $ |T_s|^2 $ cross
the value $ \omega_s $ equal to $ \omega_p^0/2 $ (see
Fig.~\ref{fig6}). The first and highest peaks in the signal-field
energy spectra $ S_s $ (see Fig.~\ref{fig7}b) correspond to the
case when both the signal and idler photons are tuned at the first
resonance near the band edge. At this frequency, localization of
the down-converted fields is maximum, and the nonlinear process is
strongest. The other two sets of peaks in the signal-field energy
spectra $ S_s $ correspond to the second and third peaks in the
intensity transmission $ |T_s|^2 $.  At these frequencies, the
nonlinear interaction is somewhat weaker due to smaller field
intensities, i.e., slightly worse localization of the
down-converted fields. The spectral FWHMs for these
frequency-degenerate emissions lye in the interval from 10~nm to
15~nm (see Fig.~\ref{fig7}a). For these values of the angle $
\theta_s $ the best available constructive interference in the
structure occurs. In a relatively narrow frequency interval the
signal and idler fields are enhanced by constructive interference.
Outside this frequency interval, destructive interference occurs.
For the other values of the angle $ \theta_s $ constructive
interference is less pronounced, the range of frequencies having
constructive interference is broader, and spectral shapes with
several peaks may occur (see Fig.~\ref{fig5}b). We thus obtain
down-converted fields with spectral FWHMs reaching even 80nm (see
Fig.~\ref{fig7}a). In this case the FWHM indicates the entire
active width, which might include several peaks.  However,
photon-pair generation rates are low in this case.

\begin{figure}    
 {\raisebox{4 cm}{a)} \hspace{0mm}
 \resizebox{0.9\hsize}{!}{\includegraphics{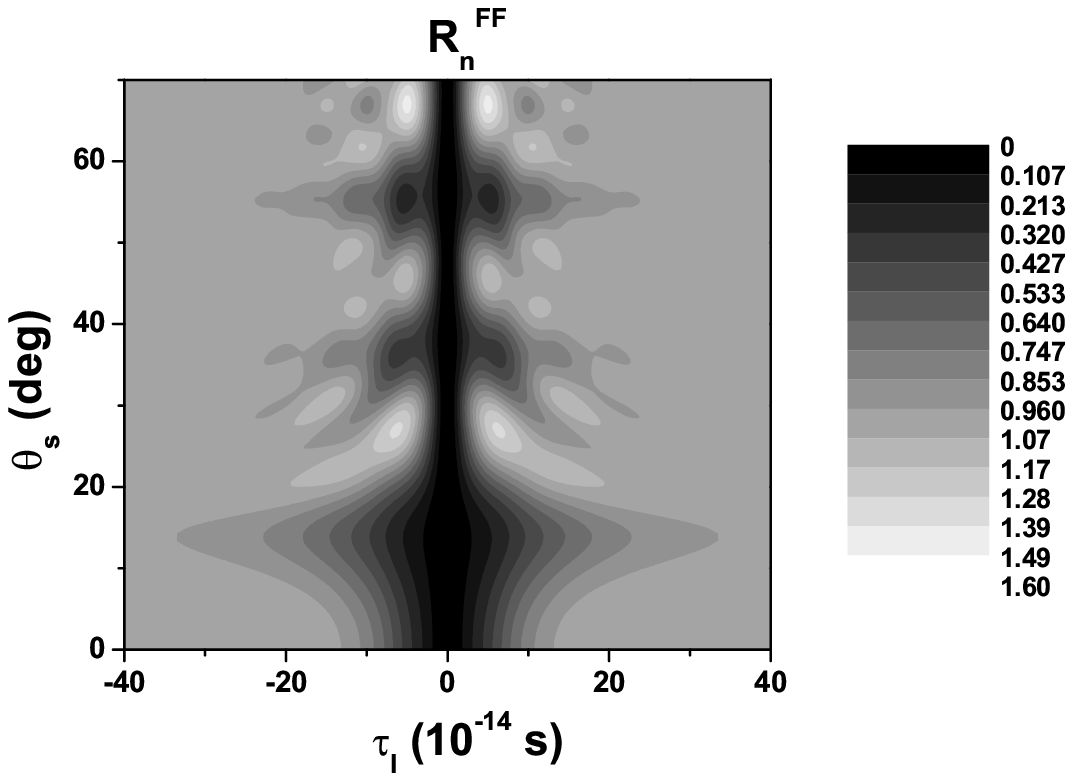}}

 \vspace{5mm}
 \hspace{-20mm} \raisebox{2 cm}{b)} \hspace{5mm}
 \resizebox{0.6\hsize}{0.3\hsize}{\includegraphics{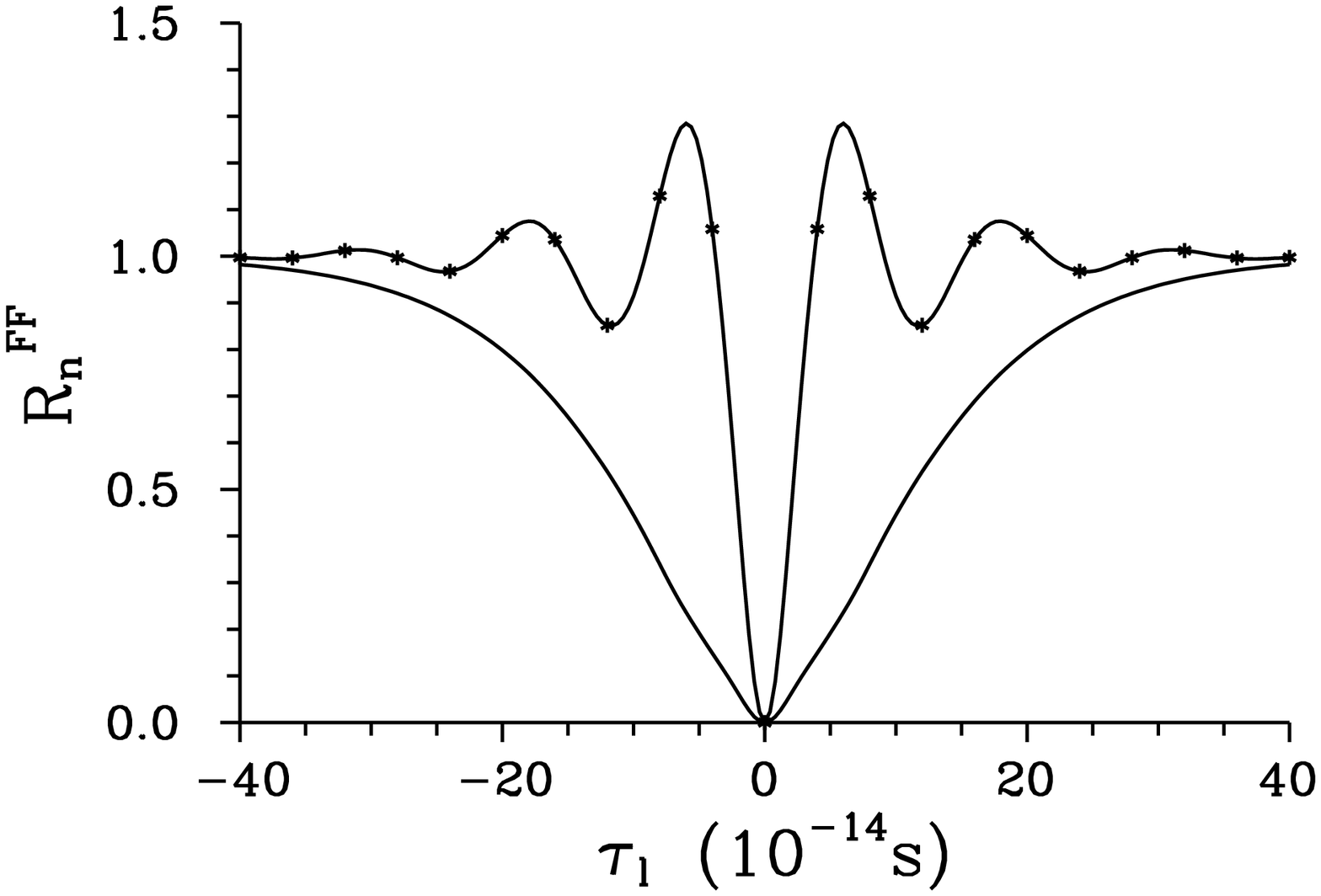}}}
 \vspace{2mm}

 \caption{(a) Normalized coincidence-count rate $ R_n^{FF} $ of mode $
 FF $ in Hong-Ou-Mandel interferometer
 as a function of relative time delay $ \tau_l $ and angle
 $ \theta_s $ of emission of a signal photon;
 (b) cuts along the lines $ \theta_s = 14 $~deg (solid curve) and
 $ \theta_s = 28 $~deg (solid curve with *); cw pumping is assumed.}
\label{fig8}
\end{figure}

Quantum correlations (entanglement) between the signal and idler
photons in a photon pair are visible in the profile of the
normalized coincidence-count rate $ R_n(\tau_l) $ in a
Hong-Ou-Mandel interferometer (see Fig.~\ref{fig8} for mode $ FF
$).
For values of the angle $ \theta_s $ with strong constructive
interference ($ \theta_s \approx 14, 35, 55 $~deg) this profile is
given by a relatively broad dip (see Fig.~\ref{fig8}b). On the
other hand, oscillating curves with several local minima and
maxima characterize profiles of the normalized coincidence-count
rates $ R_n(\tau_l) $ for other values of the angle $ \theta_s $.
A typical profile for this case is shown in Fig.~\ref{fig8}b.
These profiles reflect broader spectra of the down-converted
fields along these angles as well as a nonzero difference of the
central frequencies $ \omega_s^c $ and $ \omega_i^c $ of the
signal and idler fields. Period of these oscillations is
proportional to $ 1/(\omega_s^c-\omega_i^c) $.

\begin{figure}    
 {\raisebox{4 cm}{a)} \hspace{5mm}
 \resizebox{0.7\hsize}{!}{\includegraphics{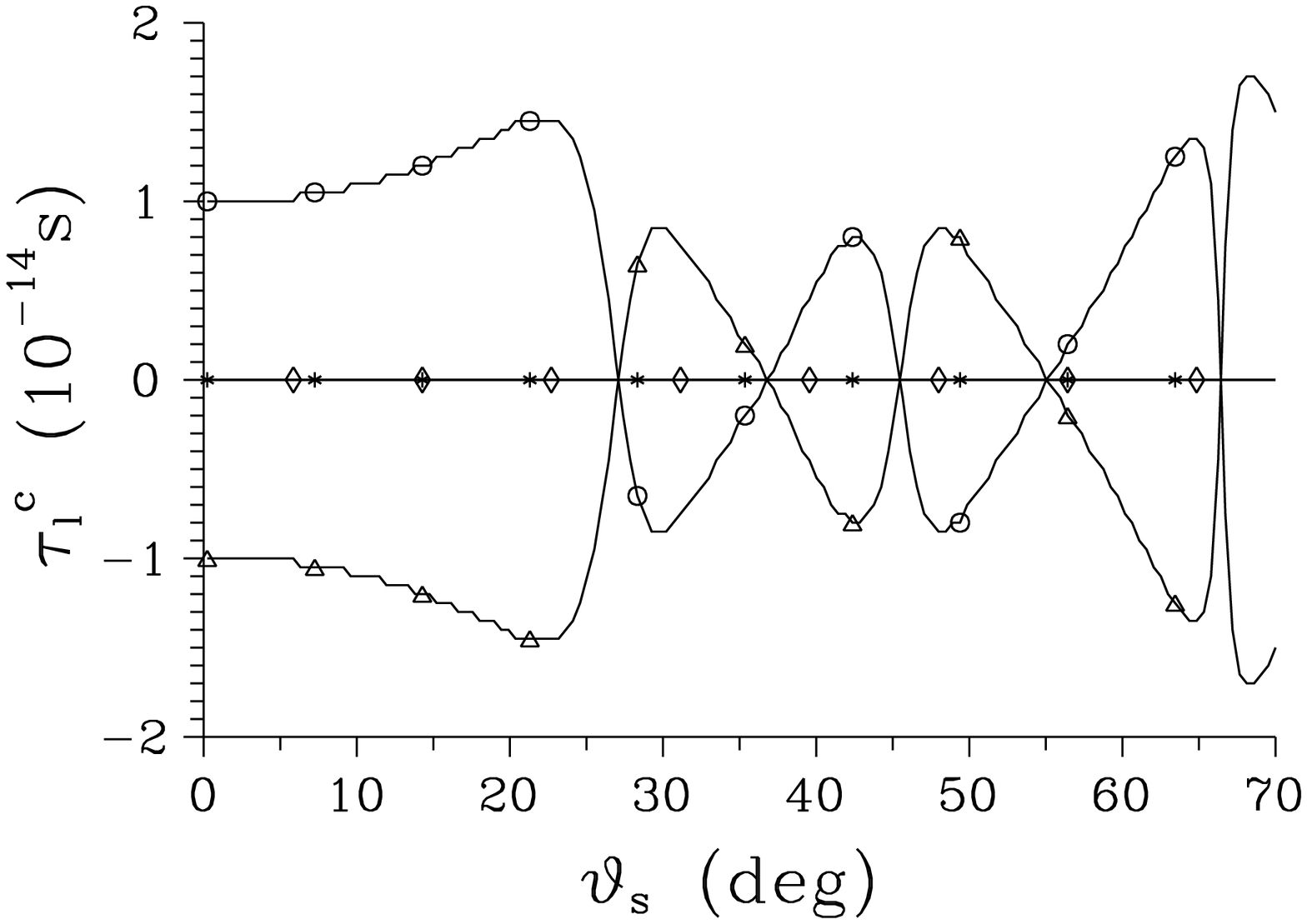}}

 \vspace{5mm}
 \raisebox{4 cm}{b)} \hspace{5mm}
 \resizebox{0.7\hsize}{!}{\includegraphics{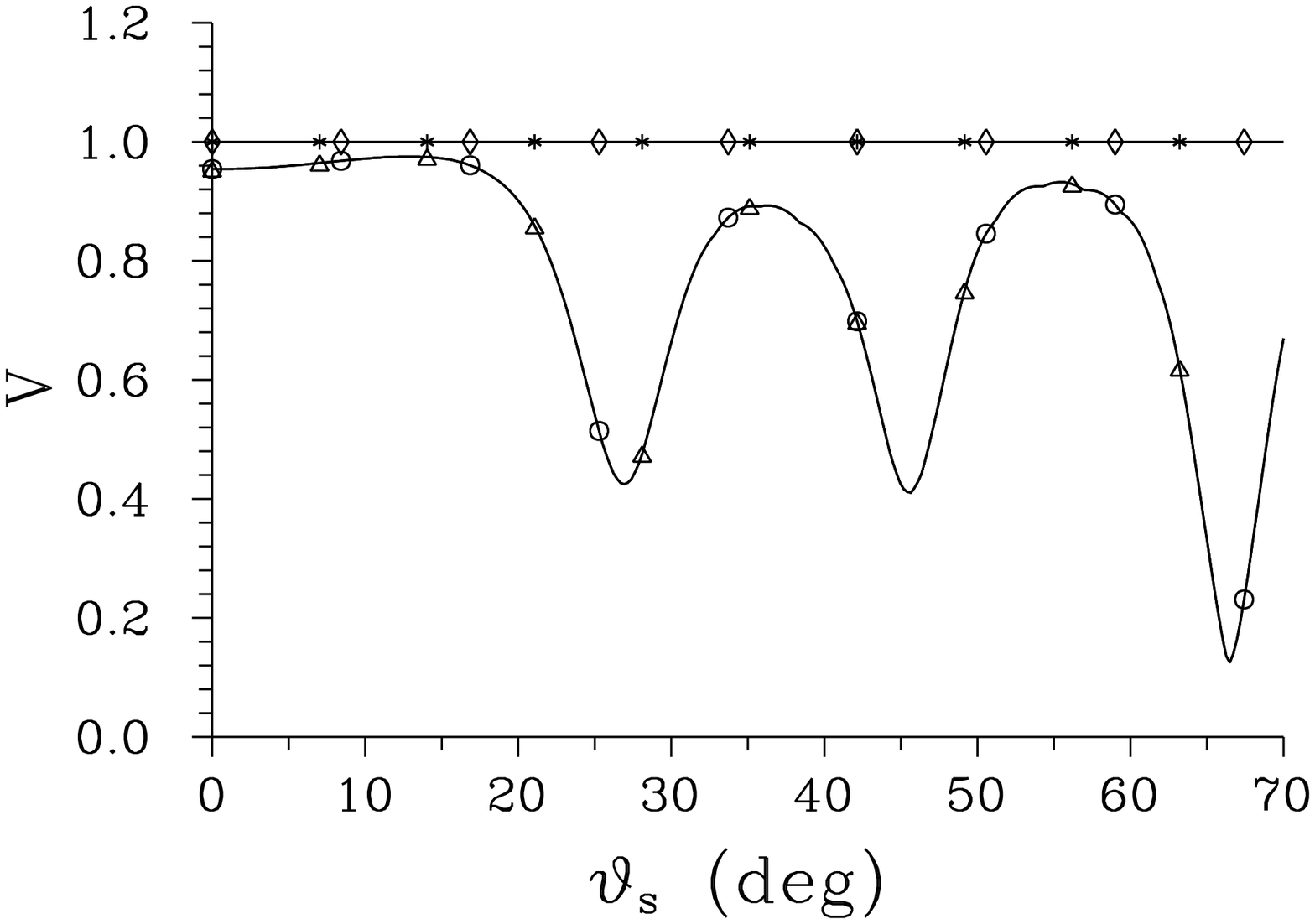}}

 \vspace{5mm}
 \raisebox{4 cm}{c)} \hspace{5mm}
 \resizebox{0.7\hsize}{!}{\includegraphics{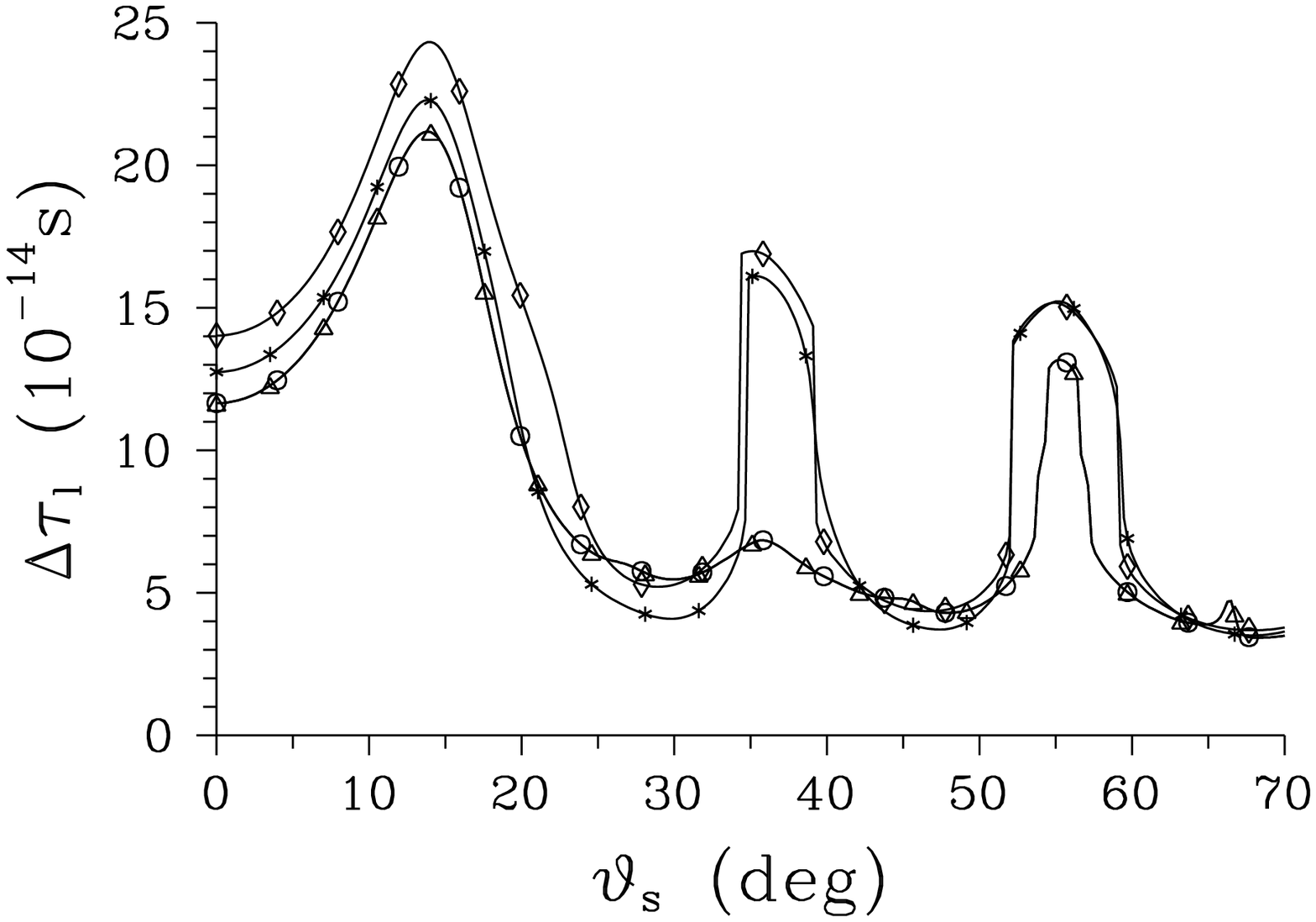}}}
 \vspace{3mm}

 \caption{Position $ \tau_l^c $ of the central dip (a), its visibility $ V $
 [$ V = \frac{\rho}{(2-\rho)} $] (b),
 and FWHM $ \Delta\tau_l $ of the dip (c) for the normalized coincidence-count rate $
 R_n(\tau_l) $ in Hong-Ou-Mandel interferometer in modes $ FF $ ($ * $), $ FB $
 ($ \triangle $), $ BF $ ($ \circ $),
 and $ BB $ ($ \diamond $) as functions of the angle $
 \theta_s $ of emission of a signal photon;
 cw pumping is assumed.}
\label{fig9}
\end{figure}

Position $ \tau_l^c $ of the central dip, its FWHM $ \Delta\tau_l
$, and visibility $ V $ of the normalized coincidence-count rate $
R_n $ as functions of the angle $ \theta_s $ are shown in
Fig.~\ref{fig9}.
We can see in Fig.~\ref{fig9}a that the maximum overlap of the
signal and idler photon fields is reached for a nonzero mutual
time delay $ \tau_l $ for nonsymmetric modes $ FB $ and $ BF $.
This means that the signal and idler photons exit the structure
with a small mutual time delay. This nonzero mutual time delay
means that interference effects inside the structure prefer the
generation of photon pairs from nonlinear layers positioned closer
to one edge of the structure, or in any case approximately
co-located. The visibility $ V $ of the coincidence-count rate $
R_n $ equal to one is naturally observed for symmetric modes $ FF
$ and $ BB $. Values of the visibility $ V $ better than 0.9 can
also be reached in nonsymmetric modes $ FB $ and $ BF $ provided
that the fields are generated along the angles $ \theta_s $ in the
vicinity of those with a strong constructive interference. The
structure of quantum correlations between the signal and idler
fields in time domain is complex and prevents the visibility $ V $
from reaching higher values along the angles $ \theta_s $, where
there are broad spectra of the down-converted fields. The width $
\Delta\tau_l $ of the dip in the normalized coincidence-count rate
$ R_n(\tau_l) $ determines the entanglement time, i.e. the time
interval in which both photons can be detected. The FWHM $
\Delta\tau_l $ of the central dip for the structure ranges from 50
fs for values of the angle $ \theta_s $ with broad spectra of the
down-converted fields, to 150-250 fs for values of the angle $
\theta_s $ at where strong constructive interference occurs (see
Fig.~\ref{fig9}c).

\subsection{Pumping by an ultrashort pulse}

\begin{figure}    
\resizebox{0.8\hsize}{!}{\includegraphics{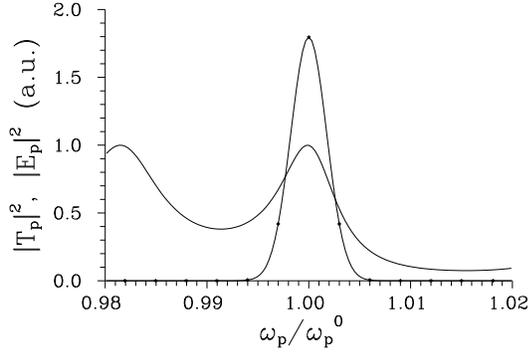}}
 \vspace{2mm}
 \caption{Pump-field intensity transmission $ |T_p|^2 $ (solid curve) and squared
 modulus $ |{\bf E}_{p_F,TE}|^2 $ of the pump-field
 amplitude spectrum (solid curve with $ * $) as
 functions of the normalized pump-field frequency $ \omega_p/\omega_p^0 $.}
\label{fig10}
\end{figure}

We now assume that the structure is pumped by an ultrashort pulse
with an unchirped Gaussian profile [see Eq.(\ref{2})] having pulse duration $
\tau_p $ equal to 200~fs, and central (carrier) frequency $ \omega_p^0 $ at the
wavelength of 664.5~nm. The squared modulus $
|E_p|^2 $ of the incident pump-field amplitude spectrum $ {\bf
E}_{p_F,TE}(\omega_p) $ fits well into the peak of resonance of
the pump-field intensity transmission $ |T_p(\omega_p)|^2 $, as
depicted in Fig.~\ref{fig10}.

\begin{figure}    
 \resizebox{0.9\hsize}{!}{\includegraphics{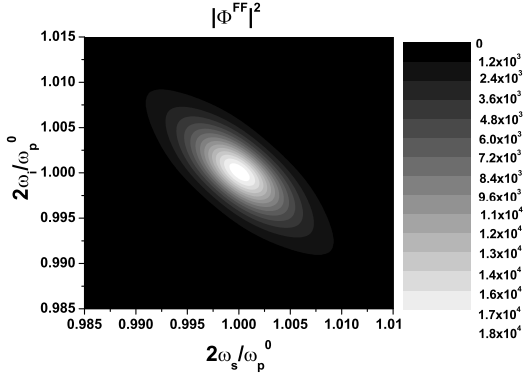}}
 \vspace{2mm}

 \caption{The probability $ |\phi^{FF}|^2 $ of
 having a photon pair in mode $ FF $ with a signal photon at the normalized frequency
 $ 2\omega_s/\omega_p^0 $ and an idler photon at the normalized frequency
 $ 2\omega_i/\omega_p^0 $ for pulsed pumping; the probability
 $ |\phi^{FF}|^2 $ is normalized such that one photon pair is emitted, i.e.
 $ 4 \int d\omega_s \int d\omega_i |\phi^{FF}(\omega_s,\omega_i)|^2
 (\omega_p^0)^2 = 1 $; $ \theta_s = 14 $~deg.}
\label{fig11}
\end{figure}

\begin{figure}    
 \resizebox{0.9\hsize}{!}{\includegraphics{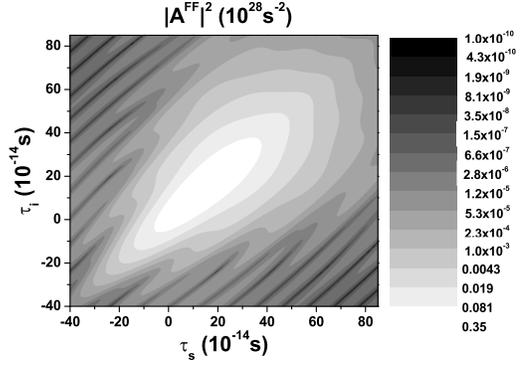}}
 \vspace{2mm}

 \caption{The probability $ |{\cal A}^{FF}|^2 $ of
 detecting a signal photon at time $ \tau_s $ and its twin idler
 photon at time $ \tau_i $ in mode $ FF $ for pulsed pumping;
 the probability $ |{\cal A}^{FF}|^2 $
 is normalized such that one photon pair is emitted, i.e.
 $ \int d\tau_s \int d\tau_i |{\cal A}^{FF}(\tau_s,\tau_i)|^2 = 1
 $; $ \theta_s = 14 $~deg; logarithmic scale is used on the  $ z $ axis.}
\label{fig12}
\end{figure}

Only signal and idler photons with frequencies $ \omega_s $ and $
\omega_i $ for which the sum $ \omega_s + \omega_i $ lyes within
the pump-pulse spectrum may be generated. The allowed values of
the frequencies $ \omega_s $ and $ \omega_i $ are indicated in the
graph of the probability $ |\phi(\omega_s,\omega_i)|^2 $ of
emitting a signal photon at frequency $ \omega_s $ and its twin at
frequency $ \omega_i $. A typical shape of the probability $
|\phi(\omega_s,\omega_i)|^2 $ is shown in Fig.~\ref{fig11} for
mode $ FF $.
We note that the shape of the probability $ |\phi
(\omega_s,\omega_i)|^2 $ in Fig.~\ref{fig11}  may be considered as being
composed of many curves defined above the lines $ \omega_s +
\omega_i = \omega_p  = {\rm const} $, and it may be assumed to be linearly
proportional to those characterizing cw pumping at the frequency $
\omega_p $. The probability $ |{\cal A}(\tau_s,\tau_i)|^2 $ of
detecting a signal photon at time $ \tau_s $ and its twin at time
$ \tau_i $ depicted in Fig.~\ref{fig12} shows that both signal and
idler photons occur in sharp time windows as a consequence of
pumping by an ultrashort pulse.
The probability $ |{\cal A}(\tau_s,\tau_i)|^2 $ has a typical
droplet shape that reflects the fact that the longer the photons
from a pair propagate inside the structure, the greater the average
difference of the occurrence times $ \tau_s $ and $ \tau_i $ of
the photons. The structure with local minima and maxima farther
from the diagonal in Fig.~\ref{fig12} reflects interference in the
layered structure.

Because the pump-pulse amplitude spectrum $ {\bf E}_{p_F,TE} $
fits well into the peak of resonance of the structure, and the
emitted frequencies of the down-converted fields have also to fit
into the structure, the energy spectra of the signal and idler fields
with pulsed pumping are very similar to those obtained for cw
pumping. Also, the behavior of photon pairs generated by pulsed pumping
in the Hong-Ou-Mandel interferometer is similar to that observed with
cw pumping; only the visibilities $ V $ of the normalized
coincidence-count rates $ R_n(\tau_l) $ are slightly worse.

\begin{figure}    
 \resizebox{0.9\hsize}{!}{\includegraphics{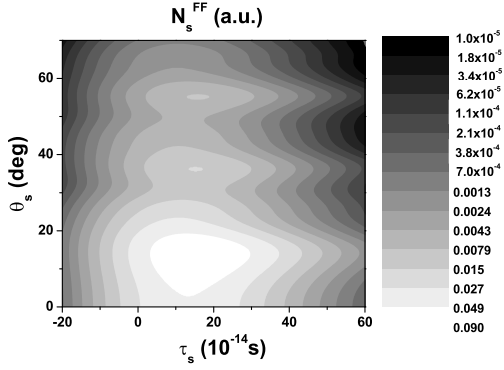}}
 \vspace{2mm}

 \caption{Photon flux $ {\cal N}_s^{FF} $ of mode $ FF $ as
 a function  of time $ \tau_s $ of detection of a signal photon and angle
 $ \theta_s $ of emission of a signal photon; pulsed pumping is assumed;
 logarithmic scale is used on the $ z $ axis.}
\label{fig13}
\end{figure}

\begin{figure}    
 {\raisebox{4 cm}{a)} \hspace{5mm}
 \resizebox{0.7\hsize}{!}{\includegraphics{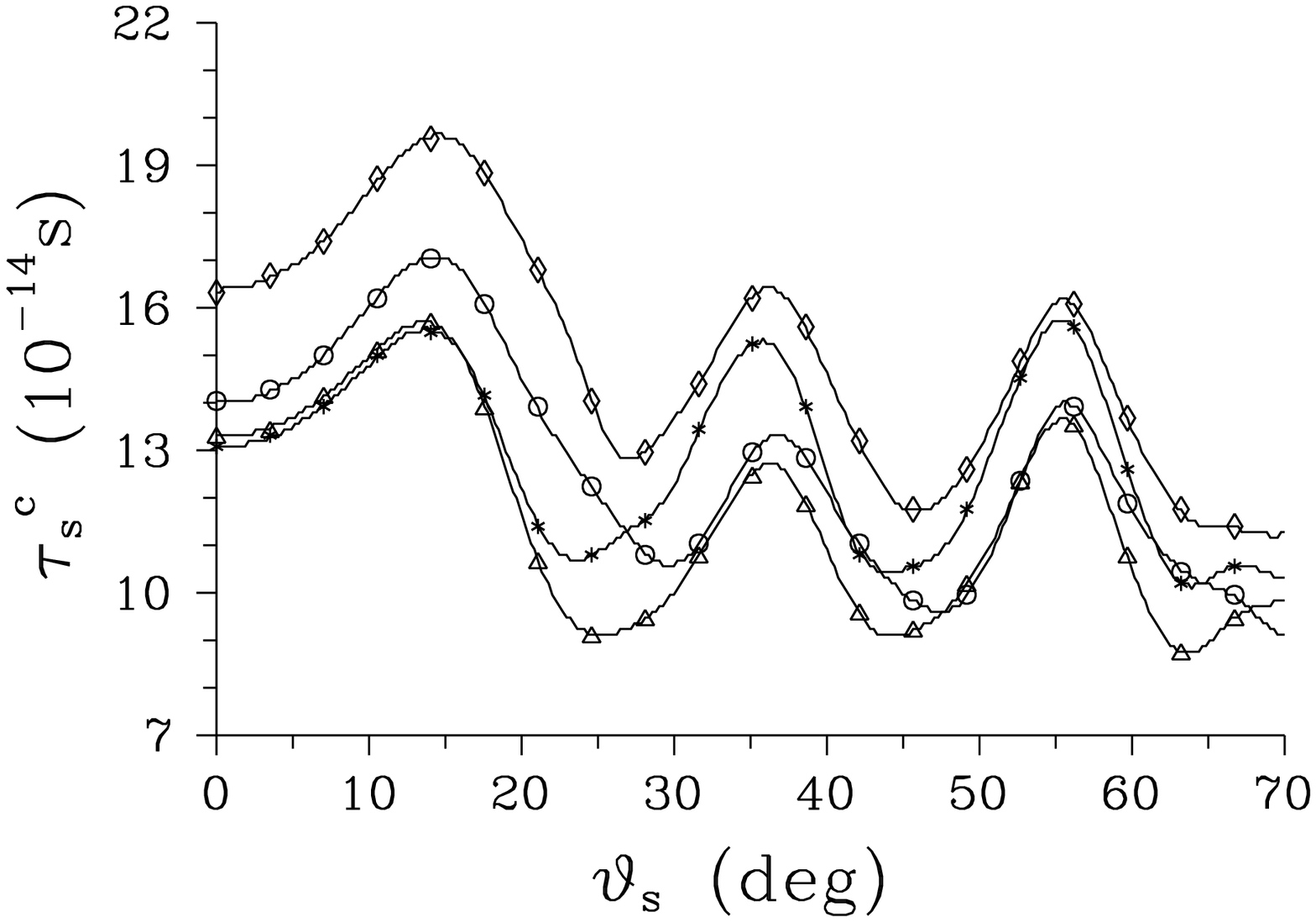}}

 \vspace{5mm}
 \raisebox{4 cm}{b)} \hspace{5mm}
 \resizebox{0.7\hsize}{!}{\includegraphics{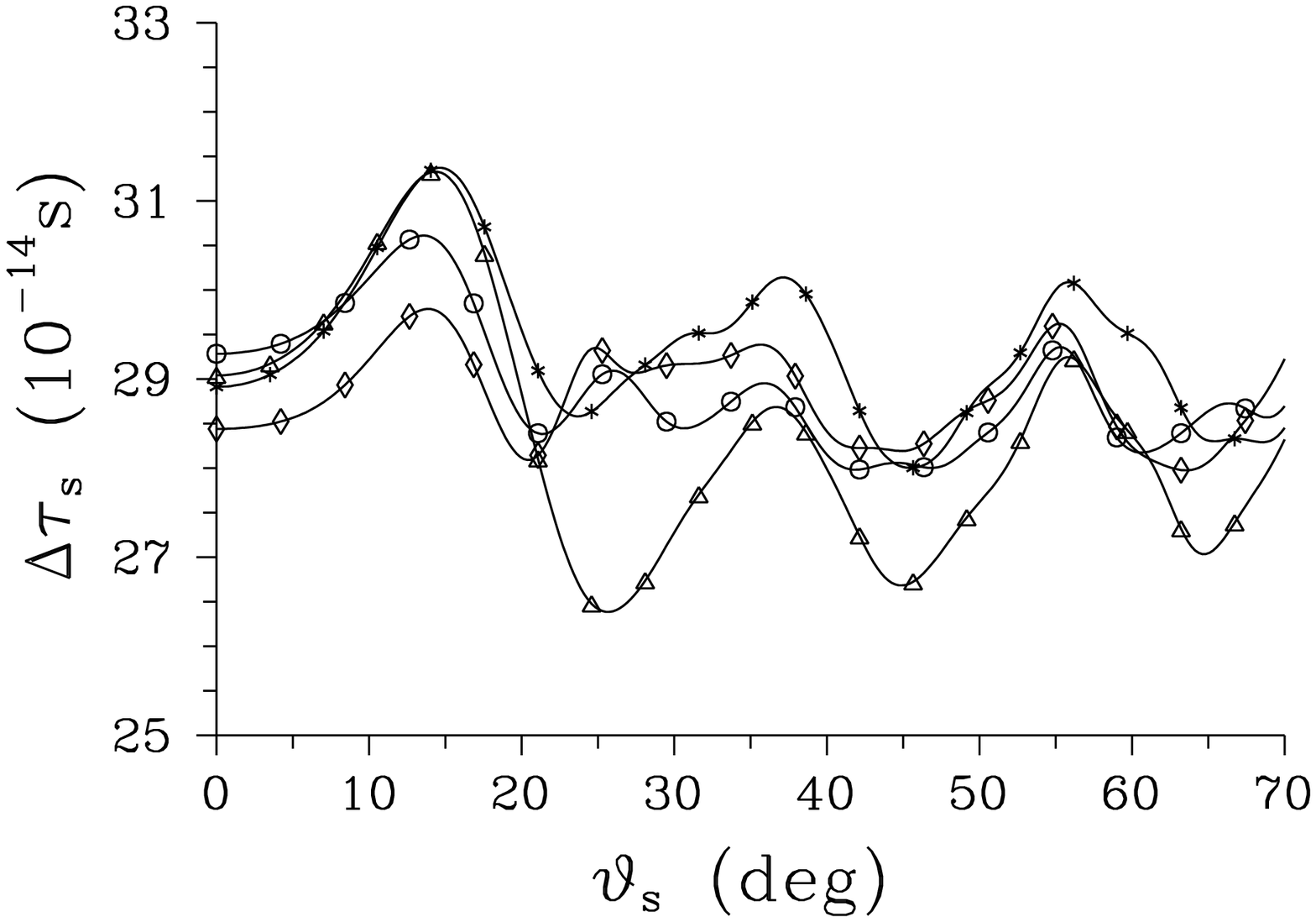}}

 \vspace{5mm}
 \raisebox{4 cm}{c)} \hspace{5mm}
 \resizebox{0.7\hsize}{!}{\includegraphics{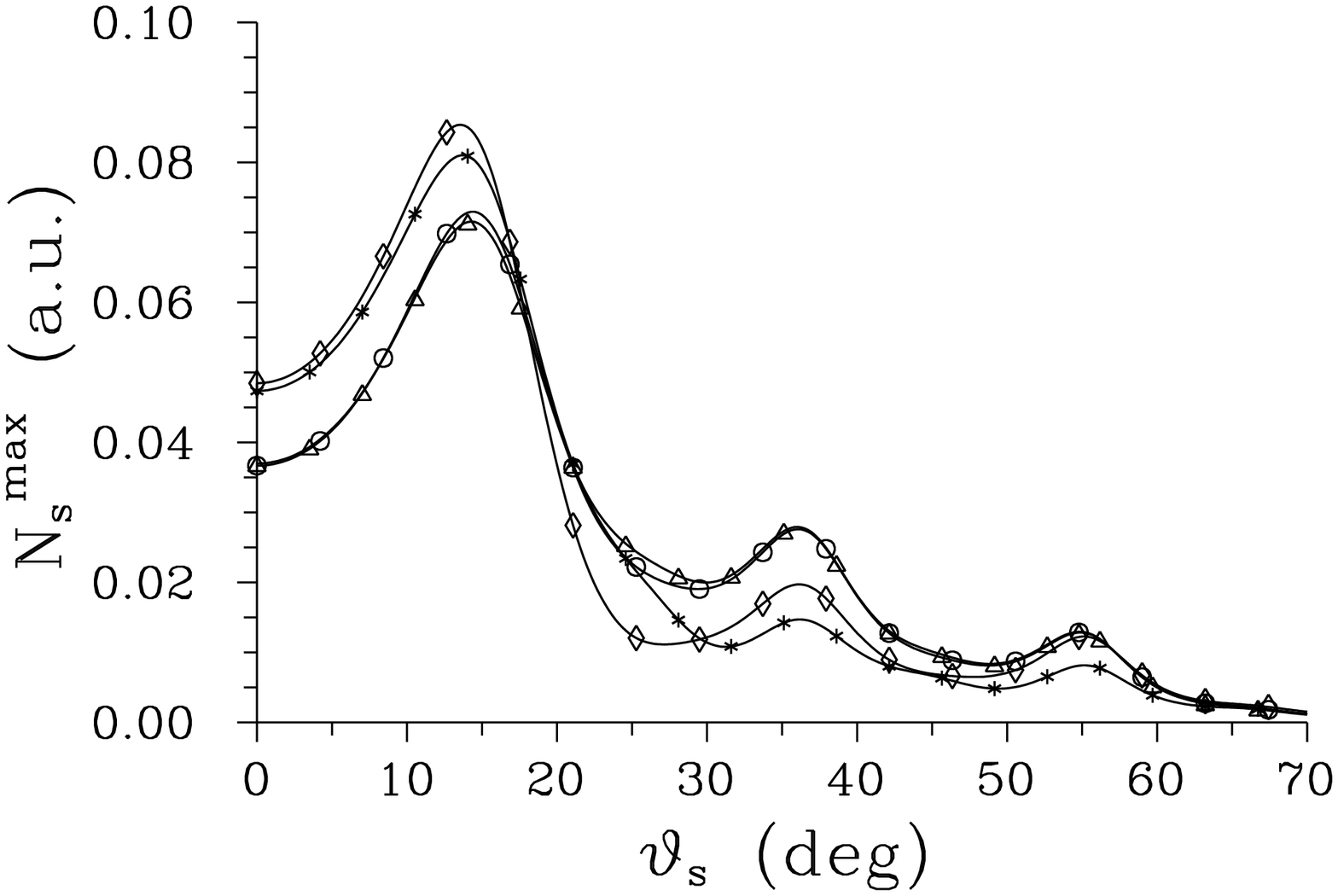}}}
 \vspace{3mm}

 \caption{Time delay $ \tau_s^c $ (a), FWHM $ \Delta\tau_s $ (b), and
  photon flux $ N_{s}^{\rm max} $ at the center ($ \tau_s = \tau_s^c $)
  for the pulsed signal field in modes $ FF $ ($ * $),
  $ FB $ ($ \triangle $), $ BF $ ($ \circ $),
  and $ BB $ ($ \diamond $) as they depend on the
  angle $ \theta_s $ of emission of a signal photon;
  pulsed pumping is assumed.}
\label{fig14}
\end{figure}

The signal and idler fields are now emitted in the form of
ultrashort, pulsed fields in multimode Fock states. A typical
dependence of the signal-field photon flux $ N_{s}^{FF} $ at time
$ \tau_s $ and angle $ \theta_s $ is shown in Fig.~\ref{fig13}.
Fig.~\ref{fig13} suggests that the photon flux $ N_{s} $ typically
spreads to longer times as a consequence of zig-zag movement, i.e.
multiple scattering, of photons inside the structure.
Time delay $ \tau_s^c $ and FWHM $ \Delta\tau_s $ of the pulsed
signal field as well as photon flux $ N_{s}^{\rm max} $ at the
center of the pulsed field as they depend on the angle $ \theta_s
$ are depicted in Fig.~\ref{fig14}.
The time delay $ \tau_s^c $ of the signal pulse with respect to
the pump pulse ranges from 70 fs to 200 fs. It reaches its maximum
values for values of the angle $ \theta_s $ with strong
constructive interference ($ \theta_s \approx 14, 35, 55 $~deg).
FWHM $ \Delta\tau_s $ of the signal pulse goes from 270~fs to
320~fs. The narrower the spectrum of the signal field (given by $
\Delta\lambda_s $ in wavelengths) the greater the signal-field
pulse duration $ \Delta\tau_s $ (compare Figs.~\ref{fig7}a and
\ref{fig14}b). The largest value of the photon flux $ N_{s}^{\max}
$ can be reached for the angle $ \theta_s $ equal to 13.8~deg (see
Fig.~\ref{fig14}c).

These results clearly show that the considered photonic-band-gap
structure maintains a pulsed character of the nonlinear process
and allows generation of pulsed down-converted fields with an
ultrashort duration. Such pulsed fields are required in many
experiments with photon pairs in which time synchronization of
several photon pairs is necessary.

\subsection{Efficiency of the nonlinear process}

The model that we have developed is one dimensional with respect
to the spatial coordinates, and does not include all aspects
(especially those related to transverse profiles of the
interacting optical fields) of the nonlinear process in a real
structure. For this reason we do not determine the absolute values
of photon-pair generation rates. Instead, we judge the efficiency
of the suggested structure (stemming from constructive
interference of the interacting optical fields) with respect to an
ideal reference structure which fully exploits the nonlinearity,
but does not rely on interference.

This reference structure has formally all linear indices of
refraction equal to one, so effects on the boundaries between
layers are suppressed. It is also assumed that the nonlinear
process is fully phase matched. The orientations of nonlinear layers,
as well as polarizations of the interacting optical fields, are
such that the greatest nonlinear effect occurs. A photon pair
emitted from this structure is described by the following output
state $ |\psi\rangle_{s,i}^{\rm ref} $ [compare Eq.~(\ref{20})]:
\begin{eqnarray}    
 |\psi\rangle_{s,i}^{\rm ref} &=& - \frac{i}{2\sqrt{2\pi}c}
  \int_{0}^{\infty} \, d\omega_p \int_{0}^{\infty} \, d\omega_s \sqrt{\omega_s}
  \sqrt{\omega_p-\omega_i}\nonumber \\
 & & \hspace{-1cm} \mbox{} |{\bf E}_{p_F}^{(+)}(0,\omega_p)| \sum_{l=1}^{N}
  \max({\bf d}^{(l)}) L_l
  \hat{a}_{s}^{\dagger}(\omega_s) \hat{a}_{i}^{\dagger}(\omega_p-\omega_s)
  |{\rm vac} \rangle, \nonumber \\
  & &
\label{50}
\end{eqnarray}
where $ \hat{a}_{s}^{\dagger}(\omega_s) $ [$
\hat{a}_{i}^{\dagger}(\omega_i) $] denotes a creation operator of
a photon in the signal [idler] field. The function $ \max $ used
in Eq.~(\ref{50}) gives the maximum value from the elements of a
tensor in its argument.

The relative photon-pair generation rate $ \eta^{mn}(\omega_s) $
of a pair with a signal photon at frequency $ \omega_s $ in mode $
mn $ is then given as:
\begin{equation}   
  \eta^{mn}(\omega_s) = \frac{S_s^{mn}(\omega_s) }{S_s^{\rm
  ref}(\omega_s) } ,
\label{51}
\end{equation}
where the signal-field energy spectrum $ S_s^{mn} $ is given in
Eq.~(\ref{35}), and the signal-field energy spectrum $ S_s^{\rm
  ref} $ of the reference structure is determined in the same way using
the output state $ |\psi\rangle_{s,i}^{\rm ref} $ written in
Eq.~(\ref{50}).

\begin{figure}    
 \resizebox{0.7\hsize}{!}{\includegraphics{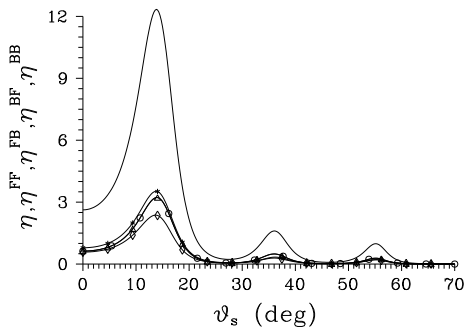}}
 \vspace{5mm}
 \caption{Relative photon-pair generation rates $ \eta^{FF} $ ($ * $), $
 \eta^{FB} $ ($ \triangle $), $ \eta^{BF} $ ($ \circ $),
 $ \eta^{BB} $ ($ \diamond $), and overall relative photon-pair generation
 rate $ \eta $ ($ \eta = \eta^{FF} + \eta^{FB} + \eta^{BF}
 + \eta^{BB} $) at the signal-field frequency $ \omega_s = \omega_p^0/2 $
 for cw pumping as a function  of angle $ \theta_s $ of emission of a signal photon.}
\label{fig15}
\end{figure}

The relative photon-pair generation rates $ \eta^{FF} $, $
\eta^{FB} $, $ \eta^{BF} $, and $ \eta^{BB} $ for the signal-field
frequency $ \omega_s = \omega_p^0/2 $ as a function of the angle
$ \theta_s $ are shown in Fig.~\ref{fig15} assuming cw pumping.
The relative photon-pair generation rates for pulsed pumping are
lower in comparison with cw pumping, because frequencies at the
edges of the pump-field spectrum have lower efficiencies than
those in the middle. Despite this, the considered pulsed pumping
has the overall relative photon-pair generation rate $ \eta $
equal to 9, whereas $ \eta = 12.5 $ for cw pumping for $ \theta_s
\approx 13.8 $~deg. The structure that we consider is approximately 12
times more efficient than the reference structure,
provided we consider all the emitted photon pairs (given by
the overall relative photon-pair generation rate $ \eta $ in
Fig.~\ref{fig15}). Even when we restrict ourselves only to  the
mode $ FF $, we still have an enhancement of the nonlinear process
by a factor of three. For comparison purposes, one layer of GaN of
thickness $ 25 \times
117 $~nm (i.e., containing the same amount of nonlinear material
as our structure) has the overall relative photon-pair
generation rate $ \eta $ equal to 0.09 for values of the angle $
\theta_s $ less than 20~deg. The inclusion of linear layers of AlN
inside the structure and the accompanying interference
effects thus increase the photon-pair generation rates by two
orders of magnitude.

\section{Conclusion}

We have developed a quantum model of spontaneous parametric
down-conversion (generating entangled photon pairs) in
one-dimensional, nonlinear, layered media. Using the model we
have determined a two-photon amplitude,
 and we have provided
measurable characteristics of the down-converted fields: marginal
signal and idler energy spectra, time-dependent photon fluxes of
the signal and idler fields, coincidence-count interference
patterns in the Hong-Ou-Mandel interferometer, and photon-pair
generation rates.

A specific structure was suggested as an efficient source of
photon pairs, and interference effects of the interacting fields can
enhance the photon-pair generation rates by as much as  two hundred
times. The widths of the spectra of the down-converted fields
and entanglement time of photons in a pair depend strongly on the
angle of emission, and vary from 10~nm to 80~nm and from 50~fs to
250~fs, respectively. Pumping the structure with an ultrashort
pulse of time duration of several hundreds of fs can lead to the generation of pulsed
signal and idler fields extending over several hundreds of fs.

Therefore, we conclude that nonlinear one-dimensional photonic-band-gap structures represent a
promising, new efficient source of photon pairs.

\acknowledgments{This work was supported by the ESF project COST
P11 (COST-STSM-P11-79), COST project OC P11.003, AVOZ10100522, and
MSM6198959213 of the Czech Ministry of Education. Support coming
from cooperation agreement between Palack\'{y} University and
University La Sapienza in Rome is acknowledged.}

\end{document}